\documentclass[sn-mathphys,Numbered]{sn-jnl}

\usepackage{graphicx}%
\usepackage{multirow}%
\usepackage{amsmath,amssymb,amsfonts}%
\usepackage{amsthm}%
\usepackage{mathrsfs}%
\usepackage[title]{appendix}%
\usepackage{xcolor}%
\usepackage{textcomp}%
\usepackage{manyfoot}%
\usepackage{booktabs}%
\usepackage{algpseudocode}%
\usepackage{listings}%

\usepackage[ruled]{algorithm2e}
\usepackage{array}
\usepackage{hyperref}

\definecolor{forestgreen}{rgb}{0.33,0.61,0.34}

\definecolor{newcolor}{HTML}{ff7f00}

\usepackage[normalem]{ulem}

\newcommand{\stkout}[1]{\ifmmode\text{\sout{\ensuremath{#1}}}\else\sout{#1}\fi}

\theoremstyle{thmstyleone}%

\theoremstyle{thmstyletwo}%

\theoremstyle{thmstylethree}%

\begin{document}

\title[Article Title]{Perturbation theory for evolution of cooperation on networks}

\author[1]{\fnm{Lingqi} \sur{Meng}}

\author*[1,2]{\fnm{Naoki} \sur{Masuda}}\email{naokimas@gmail.com}

\affil[1]{\orgdiv{Department of Mathematics}, \orgname{University at Buffalo}, \orgaddress{\street{State University of New York}, \city{Buffalo}, \postcode{14260-2900}, \state{NY}, \country{USA}}}

\affil[2]{\orgdiv{Computational and Data-Enabled Science and Engineering Program}, \orgname{University at Buffalo}, \orgaddress{\street{State University of New York}, \city{Buffalo}, \postcode{14260-5030}, \state{NY}, \country{USA}}}

\abstract{Network structure is a mechanism for promoting cooperation in social dilemma games. In the present study, we explore graph surgery, i.e., to slightly perturb the given network, towards a network that better fosters cooperation. To this end, we develop a perturbation theory to assess the change in the propensity of cooperation when we add or remove a single edge to/from the given network. Our perturbation theory is for a previously proposed random-walk-based theory that provides the threshold benefit-to-cost ratio, $(b/c)^*$, which is the value of the benefit-to-cost ratio in the donation game above which the cooperator is more likely to fixate than in a control case, for any finite networks. We find that $(b/c)^*$ decreases when we remove a single edge in a majority of cases and that our perturbation theory captures at a reasonable accuracy which edge removal makes $(b/c)^*$ small to facilitate cooperation. In contrast, $(b/c)^*$ tends to increase when we add an edge, and the perturbation theory is not good at predicting the edge addition that changes $(b/c)^*$ by a large amount. Our perturbation theory significantly reduces the computational complexity for calculating the outcome of graph surgery.}


\maketitle

\section{Introduction}

Since Darwin's time, explaining cooperative behavior in groups of self-interested individuals has been a challenge \cite{trivers1971evolution, axelrod1984evolution, hofbauer1998evolutionary, nowak2006five, nowak2006evolutionary, henrich2007humans, sigmund2010calculus, bowles2011cooperative}. Game theory including evolutionary game theory has shown that a population of self-interested individuals playing a social dilemma game of the prisoner's dilemma type does not sustain cooperation without an additional mechanism. To explain cooperation in social dilemma situations in nature including in biological populations and to promote cooperation in human society, there have been proposed various mathematical mechanisms to support cooperation. Population structure as represented by contact networks of individuals is one such mechanism. The structure of contact networks constrains who can interact with whom and promotes emergence and endurance of clusters of cooperative players in local regions in spatial lattices \cite{axelrod1984evolution, nowak1992evolutionary, nowak1993spatial, szabo2007evolutionary} and adjacent pairs of nodes in general networks \cite{santos2005scale, ohtsuki2006simple, santos2006evolutionary, szabo2007evolutionary, allen2017evolutionary}.

A major indicator of the success of a mutant trait in evolutionary dynamics is the fixation probability. It is defined as the probability that the mutant type will spread and eventually occupy the entire population as a result of evolutionary dynamics, given an initial distribution of mutants \cite{nowak2004emergence, ewens2004mathematical, lieberman2005evolutionary, nowak2006evolutionary}. When each individual is in either of the two types (i.e., wild and mutant) at any given time and the population structure is described by a network on $N$ nodes, the state of the network is specified by an $N$-dimensional binary vector of which the $i$th entry encodes the type of the $i$th node. In the absence of mutation, the fixation probability of the mutant starting from the state in which all the nodes are of the wild type is equal to $0$. The fixation probability of the mutant is equal to $1$ if all the nodes are initially mutant. For general initial conditions, the exact solution of the fixation probability requires solving a linear system of $2^N-2$ equations \cite{lieberman2005evolutionary, ohtsuki2006simple}. Therefore, it is difficult to exactly compute the fixation probability except for small networks, highly symmetric networks, or networks with other mathematically convenient properties. 

We focus on social dilemma situations, in particular the prisoner's dilemma game, in the present paper. In the prisoner's dilemma, the wild and mutant types correspond to cooperator and defector, respectively, or vice versa. The calculation of the fixation probability for the prisoner's dilemma game on networks, potentially with some additional assumptions, is usually more involved than the calculation in the case of the constant selection, in which the fitness of the wild and mutant types is fixed throughout the evolutionary dynamics. In games, the fitness of an individual generally depends on how other individuals behave, which makes setting up the linear system of $2^N-2$ equations and efficiently solving it, particularly the latter, a difficult task. Under this circumstance, weak selection is an assumption that often facilitates analytical evaluation of the fixation probability of the mutant type including in social dilemma games \cite{nowak2004emergence}. Let us write down each individual's fitness as a sum of a constant term, called the baseline fitness, and the payoff that the individual receives by playing the game. By definition, weak selection means that the payoff is small compared to the baseline fitness. Under weak selection, Ohtsuki et al. developed a pair approximation theory that enables us to analytically derive the conditions under which cooperation fixates with a larger probability than a baseline on random regular graphs, i.e., random graphs in which all nodes have the same number of neighbors \cite{ohtsuki2006simple}. Furthermore, Allen et al. extended this result to the case of arbitrary networks using coalescence times from random walk theory \cite{allen2017evolutionary}. With these methods, one can avoid dealing with a set of $2^N-2$ linear equations and calculate the leading term of the fixation probability in polynomial time in terms of $N$.

In Ref.~\cite{allen2017evolutionary}, the authors derived a key indicator to quantify the ease of cooperation in networks, i.e., the threshold benefit-to-cost ratio above which selection favors cooperation, denoted by $(b/c)^*$. In fact, substantial changes in $(b/c)^*$ may occur when one only slightly perturbs the network structure, which is an operation referred to as graph surgery \cite{allen2017evolutionary}. A carefully designed graph surgery may enhance cooperation by reducing $(b/c)^*$ by a larger amount than by a random graph surgery. For example, a small mean degree (i.e., the number of neighbors that a node has) of the network tends to induce cooperation \cite{ohtsuki2006simple, allen2017evolutionary}. Therefore, decreasing the weight of an edge or removing an edge is expected to enhance cooperation. However, this may not be an optimal choice. Which particular edge should we perturb or remove to efficiently enhance cooperation? One can answer this question by removing just one edge from the original network, calculating $(b/c)^*$ for the perturbed network, and repeating the same procedure for each different perturbation of the original network. However, this procedure may be computationally costly. Note that the method to calculate the fixation probability for cooperation in arbitrary networks, developed in Ref.~\cite{allen2017evolutionary}, is still computationally costly although its computational complexity is polynomial in $N$. 

In the current study, we develop a perturbation theory with the aim of predicting the direction and amount of the change in $(b/c)^*$ when one slightly perturbs the weight of an arbitrary single edge. We find that, for most networks, the actual change in $(b/c)^*$ when we remove an edge and the change predicted by our perturbation theory are strongly correlated, which makes it possible to propose a single edge to be removed for efficiently enhancing cooperation. However, the correlation between the result of direct numerical simulations and the perturbation theory is considerably weaker when one adds a new edge to the existing network. Therefore, our perturbation theory is not practically useful when one adds new edges. Compared to the direct numerical simulations, our perturbation theory is much faster, which allows us to compute the fixation probability under graph surgery in larger networks.

\section{Fixation of cooperation on networks under weak selection}

We assume that the graph $G$ is connected and undirected. We denote the set of nodes by $V = \{1, \ldots, N\}$, where $N$ is the number of nodes. For each pair of nodes $i, j\in V$, we denote the edge weight by $w_{ij} \geq 0$. If there is no edge between $i$ and $j$, we set $w_{ij} = 0$. We allow self-loops, i.e., positive values of $w_{ii}$~\cite{allen2017evolutionary}. The weighted degree of node $i$, denoted by $s_i = \sum_{j=1}^N w_{ij}$, also called the node strength, is the sum of the weight of the edges connected to the node. 

A discrete-time random walker is said to be simple if the walker located at node $i$ moves to one of its neighbors, denoted by $j$, in a single time step with probability proportional to $w_{ij}$, i.e., with probability $p_{ij} = w_{ij}/s_i$. Let $W = (w_{ij})$ be the $N\times N$ weighted adjacency matrix. The transition probability matrix $P = (p_{ij})$ of the simple random walk is given by $P = D^{-1}W$, where $D = \mathrm{diag}(s_1, \ldots, s_N)$, i.e., the diagonal matrix whose diagonal entries are equal to  $s_1, s_2, \ldots, s_N$. Let $\boldsymbol \pi = (\pi_1, \ldots, \pi_N)$ be the stationary probability vector of the random walk with transition probability matrix $P$, i.e., the solution of $\boldsymbol \pi P = \boldsymbol \pi$. It holds true that \cite{aldous1995reversible, masuda2017random}
\begin{equation}
\pi_i = \frac{s_i}{\sum_{\ell=1}^N s_\ell},\quad i\in \{1, \ldots, N\}.
\label{eq:pi_i}
\end{equation}

We use the donation game, which is a special case of the prisoner's dilemma game. In the donation game, which is a two-player game, one player, called the donor, decides whether or not to pay a cost $c$ $(>0)$. If the donor pays $c$, which we refer to as cooperation, then the other player, called the recipient, receives benefit $b$ $(>c)$. If the donor does not pay $c$, which we refer to as defection, then the donor does not lose anything, and the recipient does not gain anything. Therefore, the payoff matrix of the donation game for a pair of players is given by
\begin{equation}
\bordermatrix{
& \mathrm{C} & \mathrm{D}\cr
\mathrm{C} & b-c & -c\cr
\mathrm{D} & b & 0 },
\end{equation}
where C and D represent cooperation and defection, respectively, and the payoff values represent those for the row player. We assume that each player on a node participates in the game as donor and recipient half of the times each.

We assign 0 and 1 to the defector and cooperator, respectively. Then, we can represent a state of the entire network by a binary vector $\boldsymbol x =(x_1, \ldots, x_N) \in \{0, 1\}^N$. With this notation, the payoff of node $i$ averaged over all its neighbors is given by
\begin{align}
    f_i(\boldsymbol x) = -c x_i + b\sum_{j=1}^N p_{ij}x_j.
\end{align}
The reproductive rate of node $i$ in state $\boldsymbol x$ is given by
\begin{align}
    R_i (\boldsymbol x) = 1 + \eta f_i(\boldsymbol x),
\end{align}
where $\eta$ represents the strength of the selection. If $\eta=0$, the reproductive rate does not depend on the payoff matrix or the action (i.e., cooperation or defection) of any node. This case is equivalent to the so-called voter model. If $\eta \to 0$, the payoff weakly impacts the selection, and this limit is called the weak selection regime. The idea behind weak selection is that, in reality, many different factors may contribute to the overall fitness of an individual, and the game under consideration is just one such factor \cite{ohtsuki2006simple, allen2017evolutionary}.

\begin{figure}[!t]
  \centering
  \includegraphics[width=0.99\textwidth]{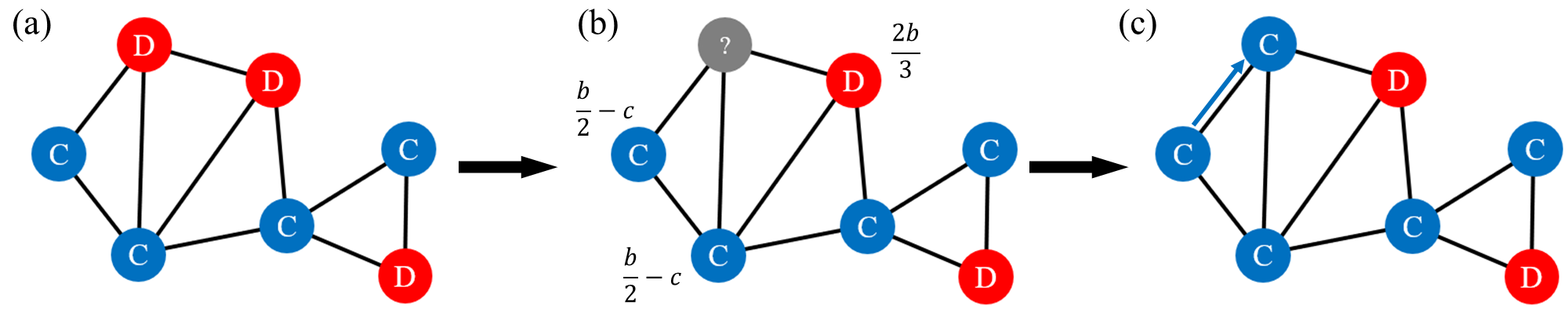}
  \caption{Death-birth process with selection on birth on the unweighted network. (a) Each individual obtains a payoff by interacting with all its neighbors. C and D represent cooperator and defector, respectively. (b) We select a node whose type to be replaced uniformly at random, shown in gray. Then, one of the three neighbors of this node, whose payoff values are indicated, will replace the gray node. We select each of the cooperating neighbors with probability $[1+\eta(b/2-c)] / [1+\eta(b/2-c) + 1+\eta(b/2-c) + 1+\eta(2b/3)] = [6+3\eta (b-2c)]/[18+2\eta (5b-6c)]$ and the defecting neighbor with probability $[1+\eta(2b/3)] / [1+\eta(b/2-c) + 1+\eta(b/2-c) + 1+\eta(2b/3)] = (3+2\eta b)/[9+\eta (5b-6c)]$ for reproduction. (c) In this example, we select the cooperating neighbor to the left, which replaces the offspring node. }
  \label{DB process}
\end{figure}

We drive evolutionary dynamics by the death-birth process with selection on birth on an arbitrary network composed of cooperators and defectors \cite{ohtsuki2006simple, allen2017evolutionary}. Specifically, we first select a node to be updated, denoted by $i$, uniformly at random. Second, we select one of the $i$'s neighbors, denoted by $j$, for reproduction with the probability proportional to $w_{ij} R_j(\boldsymbol x)$. Third, the offspring, $i$, inherits the type of $j$. This completes a single round of the evolutionary dynamics, which we schematically show in Fig.~\ref{DB process}. 

The death-birth process in any finite population without mutation will eventually reach the state in which all individuals are cooperators or defectors and halt. In other words, the cooperation or defection fixates in finite time with probability $1$. Suppose the initial condition in which one node is cooperator and the other $N-1$ nodes are defectors. There are $N$ such initial conditions depending on which node is the cooperator. We consider the initial probability distribution over all possible states that assigns probability $1/N$ to each of the states with exactly one cooperator and probability zero to all the other states. We denote by $\rho_{\mathrm{C}}$ the expectation that the cooperation fixates under this distribution of the initial state. If $\rho_{\mathrm{C}} > 1/N$, natural selection favors cooperation \cite{nowak2004emergence, ohtsuki2006simple, nowak2006evolutionary, allen2017evolutionary}. In Ref.~\cite{allen2017evolutionary}, Allen et al. showed that 
\begin{align}
    \rho_{\mathrm{C}} = \frac{1}{N} + \frac{\eta}{2N}\left[-c\tau_2 + b(\tau_3 - \tau_1)\right] + O(\eta^2),
    \label{rho_c}
\end{align}
where
\begin{align}
    \tau_k = \sum_{i=1}^N\sum_{j=1}^N \pi_i p_{ij}^{(k)}t_{ij},
    \label{tau k}
\end{align}
$p_{ij}^{(k)}$ is the $(i,j)$th entry of matrix $P^k$, which implies that $p_{ij}^{(1)} = p_{ij}$, and 
\begin{align}
    t_{ij} =
    \begin{cases} 
      0 & \mathrm{if} \ i=j, \\
      1 + \frac{1}{2}\sum_{k=1}^N(p_{ik}t_{jk} + p_{jk}t_{ik}) & \mathrm{otherwise}.
   \end{cases}
   \label{t_ij}
\end{align}
Equation (\ref{t_ij}) implies that $t_{ij} = t_{ji}$ is the mean coalescence time of two random walkers when one walker is initially located at node $i$ and the other at node $j$. Note that $p_{ij}^{(k)}$ is the $k$-step transition probability of the random walk from node $i$ to node $j$. Therefore, $\tau_k$ is the expected value of $t_{ij}$ when $i$ and $j$ are the two ends of a $k$-step random walk trajectory on $G$ under the stationary distribution \cite{allen2017evolutionary}. Equation~(\ref{rho_c}) implies that the threshold value of the benefit-to-cost ratio above which the natural selection favors cooperation (i.e., $ \rho_{\mathrm{C}} > 1/N$) is given by
\begin{align}
    \left(\frac{b}{c}\right)^* = \frac{\tau_2}{\tau_3 - \tau_1}.
    \label{bcthreshold}
\end{align}
Natural selection favors cooperation if $b/c > (b/c)^*$. 

For example, if the underlying network is regular with degree $k$, we have
\begin{align}
    &\tau_1 = N - 1,\\
    &\tau_2 = N - 2,
\end{align}
and
\begin{align}
    \tau_3 = N + \frac{N}{k} - 3,
\end{align}
such that 
\begin{align}
    \left(\frac{b}{c}\right)^* = k
\end{align}
as $N\to\infty$ \cite{allen2017evolutionary}. Note that the right-hand side of Eq. (\ref{bcthreshold}) only depends on the adjacency matrix of the network. In other words, the structure of the contact network determines whether and how much natural selection favors cooperation.

\section{Perturbation theory for graph surgery}

In this section, we develop a perturbation theory to determine the change in $(b/c)^*$ when one perturbs the weight of a single edge. To this end, we start by rewriting Eq.~(\ref{tau k}) in terms of matrices and vectors. Let $\boldsymbol 1 = (1, \ldots , 1)^\top$, where $^\top$ represents the transposition. Let $T = (t_{ij})$ be the $N\times N$ matrix of the mean coalescence time. Using these notations, we rewrite Eq.~(\ref{tau k}) as
\begin{align}
    \tau_k = \boldsymbol\pi \left( P^k\circ T \right)\boldsymbol 1, 
    \label{tau k matrix}
\end{align}
where $k=1, 2, 3$, and $\circ$ represents the Hadamard product. 

If one changes the weight of an edge $(i_0, j_0)$ by $\varepsilon$, where $| \varepsilon | \ll 1$, including the case in which we create a new edge with weight $\varepsilon$ $(> 0)$, the perturbed network remains connected and undirected. Therefore, one can still use Eq.~(\ref{bcthreshold}) to compute $(b/c)^*$. Equation~(\ref{bcthreshold}) uses Eq.~(\ref{tau k}), which requires $\boldsymbol\pi$, $P$, and $T$. We denote these variables after the perturbation by $\boldsymbol\pi(\varepsilon)$, $P(\varepsilon)$, and $T(\varepsilon)$. To distinguish the quantities before and after the perturbation, we denote these variables before the perturbation by $\boldsymbol \pi(0)$, $P(0)$, and $T(0)$.

For writing down $\boldsymbol\pi(\varepsilon)$, we denote by 
\begin{align}
    S = \sum_{i=1}^N s_{i} = \sum_{i=1}^N \sum_{j=1}^N w_{ij}
    \label{network weight}
\end{align}
the sum of the weighted degree of over all the nodes. Under a small perturbation, we carry out Taylor expansion of Eq.~\eqref{eq:pi_i} to obtain
\begin{align}
   \boldsymbol\pi(\varepsilon) = \boldsymbol\pi(0) + \varepsilon \Delta \boldsymbol\pi + o(\varepsilon), 
\end{align}
where $\Delta \boldsymbol \pi = (\Delta \pi_1, \ldots, \Delta \pi_N)$. We obtain
\begin{align}
    \Delta \pi_i = \frac{\delta_{ii_0} + \delta_{ij_0}}{S} - \frac{2\pi_i(0)}{S}, 
    \label{eq_delta_pi}
\end{align}
where $\delta_{ij}$ is the Kronecker delta. We present the derivation of Eq.~\eqref{eq_delta_pi} in Appendix~\ref{sec:derivation-Delta_pi}. 

To calculate $P(\varepsilon)$, we define a symmetric indicator function, denoted by $\chi_{i_0 j_0}$, by
\begin{align}
	\chi_{i_0 j_0}(i, j) = 
    \begin{cases} 
      1 & \mathrm{if\ } (i, j) = (i_0, j_0)\ \mathrm{or}\ (i, j) = (j_0, i_0),\\
      0 & \mathrm{otherwise}.
   \end{cases}
   \label{chi}
\end{align}
We obtain
\begin{align}
    P(\varepsilon) & = P(0) + \varepsilon \Theta^{(1)} + o(\varepsilon), \\
    P^2(\varepsilon) & = P^2(0) + \varepsilon \left[\Theta^{(1)} P(0) + P(0) \Theta^{(1)}\right] + o(\varepsilon) \nonumber\\
    & := P^2(0) + \varepsilon \Theta^{(2)} + o(\varepsilon),\\
    P^3(\varepsilon) & = P^3(0) + \varepsilon \left[\Theta^{(1)} P^2(0) + P(0) \Theta^{(1)} P(0) + P^2(0) \Theta^{(1)}\right] + o(\varepsilon) \nonumber\\
    & := P^3(0) + \varepsilon \Theta^{(3)} + o(\varepsilon),
\end{align}
where $\Theta^{(1)} = (\theta^{(1)}_{ij})$, $\Theta^{(2)} = (\theta^{(2)}_{ij})$, and $\Theta^{(3)} = (\theta^{(3)}_{ij})$ are $N \times N$ matrices whose entries are given by
\begin{align}
    \theta^{(1)}_{ij} & = \frac{\chi_{i_0 j_0}(i, j)}{s_i} - p_{ij}(0) \frac{\delta_{ii_0} + \delta_{ij_0}}{s_i}, \phantom{----------------}
    \label{theta1}
\end{align}
\begin{align}
\theta^{(2)}_{ij} & = \frac{\delta_{ii_0}p_{j_0j}(0)}{s_i} + \frac{\delta_{ij_0}p_{i_0j}(0)}{s_i} - p_{ij}^{(2)}(0)\frac{\delta_{ii_0}+\delta_{ij_0}}{s_i} \nonumber\\
& \phantom{=} + \frac{\delta_{jj_0}p_{ii_0}(0)}{s_{i_0}} + \frac{\delta_{ji_0}p_{ij_0}(0)}{s_{j_0}} - p_{ii_0}(0)p_{i_0j}(0) \frac{1}{s_{i_0}} - p_{ij_0}(0)p_{j_0j}(0) \frac{1}{s_{j_0}},
\label{theta2}
\end{align}
and
\begin{align}
    \theta^{(3)}_{ij} & = \frac{\delta_{ii_0}}{s_{i_0}}p^{(2)}_{j_0j}(0) + \frac{\delta_{ij_0}}{s_{j_0}}p^{(2)}_{i_0j}(0) - p^{(3)}_{ij}(0)\frac{\delta_{ii_0}+\delta_{ij_0}}{s_i} \nonumber \\
    & \phantom{=} + \frac{p_{ii_0}(0)p_{j_0j}(0)}{s_{i_0}} + \frac{p_{ij_0}(0)p_{i_0j}(0)}{s_{j_0}} - \frac{p_{ii_0}(0)p_{i_0j}^{(2)}(0)}{s_{i_0}} - \frac{p_{ij_0}(0)p_{j_0j}^{(2)}(0)}{s_{j_0}} \nonumber \\
    & \phantom{=} + \frac{\delta_{jj_0}}{s_{i_0}}p^{(2)}_{ii_0}(0) + \frac{\delta_{ji_0}}{s_{j_0}}p^{(2)}_{ij_0}(0) - p^{(2)}_{ii_0}(0)p_{i_0j}(0) \frac{1}{s_{i_0}} - p^{(2)}_{ij_0}(0)p_{j_0j}(0) \frac{1}{s_{j_0}}.
    \label{theta3}
\end{align}
We show the derivation of Eqs.~\eqref{theta1}, \eqref{theta2}, and \eqref{theta3} in Appendix~\ref{sec:derivation thetas}.

We next calculate $T(\varepsilon)$. Matrix $T(0) = (t_{ij}(0))$ satisfies
\begin{align}
	t_{ij}(0) = 
    \begin{cases} 
      0 & \mathrm{if}\ i = j, \\
      1 + \frac{1}{2} \left[\sum_{k=1}^{j-1} p_{ik}(0)t_{kj}(0) + \sum_{k=j+1}^{N} p_{ik}(0)t_{jk}(0) \right. \\
      \left. + \sum_{k=1}^{i-1} p_{jk}(0)t_{ki}(0) + \sum_{k=i+1}^{N} p_{jk}(0)t_{ik}(0) \right] & \mathrm{if}\ i < j, \\
      t_{ji}(0) & \mathrm{if}\ i > j,
   \end{cases}
   \label{coalesence time}
\end{align}
which we obtain by applying $t_{ij}(0) = t_{ji}(0)$ to Eq.~(\ref{t_ij}). Note that $\{p_{11}(0), p_{12}(0), \ldots, p_{NN}(0) \}$ are known from the network structure and that $\{t_{11}(0), t_{12}(0), \ldots, t_{NN}(0) \}$ are unknowns. We stack Eq.~(\ref{coalesence time}) for the different $i$ and $j$ values in lexicographical order of $(i, j)$ on the left-hand side. In other words, the first equation is $t_{11}(0) = 0$, the second equation is $t_{12}(0) - \frac{1}{2} p_{11}(0)t_{12}(0) - \frac{1}{2}\sum_{k=3}^{N} p_{1k}(0)t_{2k}(0) - \frac{1}{2} \sum_{k=2}^{N} p_{2k}(0)t_{1k}(0) = 1$, the third equation is $t_{13}(0) - \frac{1}{2} p_{11}(0)t_{13}(0) - \frac{1}{2} p_{12}(0)t_{23}(0) - \frac{1}{2} \sum_{k=4}^{N} p_{1k}(0)t_{3k}(0) - \frac{1}{2} \sum_{k=2}^{N} p_{3k}(0)t_{1k}(0) = 1$, and so on. Denote by $\mathrm{vec}(T(0))$ the thus obtained vectorization of matrix $T(0)$, i.e.,
\begin{align}
    \mathrm{vec}(T(0)) = (t_{11}(0), \ldots , t_{1N}(0); t_{21}(0), \ldots, t_{2N}(0); \ldots, t_{N1}(0), \ldots, t_{NN}(0))^\top.
    \label{vecT0}
\end{align}
Equation~(\ref{vecT0}) is a redundant expression because $T(0)$ is a symmetric matrix and its diagonal elements are equal to 0. However, we use Eq.~\eqref{vecT0} in the following text because it makes the theoretical derivations and computational implementation easier than the most compact vector form of $T(0)$, which would be $N(N-1)/2$-dimensional. Using Eq.~(\ref{vecT0}), we rewrite Eq.~(\ref{coalesence time}) as
\begin{align}
    M(0)\mathrm{vec}(T(0)) = \boldsymbol d,
\end{align}
where $M(0)$ is the $N^2\times N^2$ matrix whose entries are determined by Eq.~(\ref{coalesence time}), and $\boldsymbol d$ is the $N^2$-dimensional column vector whose $((k-1)N+k)$th entry is equal to 0 for all $k\in\{1,\ldots N\}$, and all the other entries are equal to 1. Because it also holds true that $M(\varepsilon) \mathrm{vec}(T(\varepsilon)) = \boldsymbol d$, the calculation of $T(\varepsilon)$ requires $M(\varepsilon)$, which is the matrix with perturbation, defined similarly to $M(0)$. We obtain the entries of $M(\varepsilon)$ by those of $M(0)$ with each $p_{ij}(0)$ (with $i, j \in \{1, \ldots, N\}$) being replaced by $p_{ij}(\varepsilon)$. We write the Taylor expansion of $M(\varepsilon)$ as
\begin{align}
    M(\varepsilon) = M(0) + \varepsilon \Delta M + o(\varepsilon)
    \label{M varepsilon}
\end{align}
and calculate $\Delta M$ as follows.

We write $\Delta M$ as a block matrix
\begin{align}
\Delta M = 
    \begin{pmatrix}
    \Delta_{11} & \Delta_{12} & \cdots & \Delta_{1N}\\
    \Delta_{21} & \Delta_{22} & \cdots & \Delta_{2N}\\
    \vdots & \vdots & \ddots & \vdots\\
    \Delta_{N1} & \Delta_{N2} & \cdots & \Delta_{NN}
    \end{pmatrix},
    \label{delta_M}
\end{align}
where each $\Delta_{ij}$ is an $N\times N$ matrix. We show the derivation of each $\Delta_{ij}$ in Appendix~\ref{sec:derivation-DeltaM}. We point out that the number of nonzero rows of $\Delta M$ is equal to $2\times (N-2) + (N-1) \times 2 = 4N-6$, which is much smaller than $N^2$ for a large $N$ (see Appendix~\ref{sec:derivation-DeltaM}).

To derive the first-order term of $T(\varepsilon)$ from $\Delta M$, we use Eq.~(\ref{M varepsilon}) to obtain
\begin{align}
    \mathrm{vec}(T(\varepsilon)) & = M(\varepsilon)^{-1} \boldsymbol d \nonumber\\
    & = (M(0) + \Delta M)^{-1} \boldsymbol d \nonumber\\
    & = \left[M(0) (I + \varepsilon M(0)^{-1} \Delta M + o(\varepsilon)) \right]^{-1} \boldsymbol d \nonumber\\
    & = \left[I - \varepsilon M(0)^{-1} \Delta M + o(\varepsilon) \right] M(0)^{-1} \boldsymbol d \nonumber\\
    & = \left( I - \varepsilon M(0)^{-1} \Delta M \right) \mathrm{vec}(T(0)) + o(\varepsilon) \nonumber\\
    & = \mathrm{vec}(T(0)) - \varepsilon M(0)^{-1} \Delta M \mathrm{vec}(T(0)) + o(\varepsilon).
\end{align}
Therefore, we obtain
\begin{align}
    T(\varepsilon) := T(0) + \varepsilon \Delta T + o(\varepsilon),
\end{align}
where $\Delta T$ is the $N \times N$ matrix satisfying 
\begin{align}
    \mathrm{vec}(\Delta T) = - M(0)^{-1} \Delta M \mathrm{vec}(T(0)).
\end{align}
Finally, using Eq.~(\ref{tau k matrix}), we derive the perturbed $\tau_k(\varepsilon)$ as follows:
\begin{align}
    \tau_k(\varepsilon) = \tau_k(0) + \varepsilon \Gamma_k + o(\varepsilon),
    \label{tau k epsilon}
\end{align}
where 
\begin{align}
    \Gamma_k = \Delta \boldsymbol\pi (P^k(0) \circ T(0)) + \boldsymbol \pi (0) (\Theta^{(k)} \circ T(0)) + \boldsymbol \pi (0) (P^k(0) \circ \Delta T).
\end{align}
By substituting Eq.~(\ref{tau k epsilon}) in Eq.~(\ref{bcthreshold}), we obtain
\begin{align}
    \left(\frac{b}{c}\right)^*(\varepsilon) & := \left(\frac{b}{c}\right)^*(0) + \varepsilon \Delta \left(\frac{b}{c}\right)^* + o(\varepsilon),
    \label{bcthreshold epsilon}
\end{align}
where
\begin{align}
    \Delta \left(\frac{b}{c}\right)^* = \frac{(\tau_3(0)-\tau_1(0))\Gamma_2-\tau_2(0)(\Gamma_3-\Gamma_1)}{(\tau_3(0)-\tau_1(0))^2}.
\end{align}

\section{Time complexity}
\label{algorithm}

To calculate $(b/c)^*$ for a network with $N$ nodes, the original algorithm requires calculating the mean coalescence time by solving a linear system of $N(N-1)/2$ variables, i.e., $t_{ij}$ (with $i, j \in \{1, \ldots, N \}$ and $i < j)$, which has a time complexity of $O(N^6)$. With the Coppersmith-Winograd algorithm \cite{coppersmith1987matrix}, the time complexity is reduced to $O(N^{4.75})$ \cite{allen2017evolutionary}. To determine the single edge whose removal decreases $(b/c)^*$ by the largest amount, for example, one needs to repeat this procedure for each edge. Therefore, the entire procedure with an ordinary algorithm and the Coppersmith-Winograd algorithm requires $O(N^6|E|)$ and $O(N^{4.75}|E|)$ time, respectively, where $|E|$ is the number of edges. For a sparse network, for which $|E| = O(N)$, the time complexity is $O(N^7)$ and $O(N^{5.75})$, respectively.

The matrix $\Delta M$ defined by Eq.~\eqref{delta_M} is sparse and has a special pattern. If the $i$th row of $\Delta M$ is a zero row, then the $i$th element of vector $\Delta M \mathrm{vec}(T(0))$ is zero, and we do not need to calculate it. Therefore, to calculate $\Delta M \mathrm{vec}(T(0))$, we only need to focus on its $((i_0-1)N+k)$th entries, where $k\in \{1, \ldots, N\} \setminus \{i_0\}$, $((j_0-1)N+k)$th  entries, where $k \in \{1, \ldots, N \} \setminus \{j_0 \}$, and $((k-1)N+i_0)$th and $((k-1)N+j_0)$th entries, where $k\in \{1,\ldots,N\} \setminus \{i_0, j_0\}$. All the other entries of $\Delta M \mathrm{vec}(T(0))$ are equal to $0$. We show a pseudo algorithm to calculate $\Delta T$ in Algorithm 1.

\begin{algorithm}[!t]
\DontPrintSemicolon
\SetAlgoLined
\SetNoFillComment
\LinesNotNumbered 
\caption{Pseudoalgorithm to compute $\Delta T$. Let $\left( M(0)^{-1} \right)_i$ be the $i$th column of $M(0)^{-1}$. Let $\Theta_i$ be the $i$th row of $\Theta^{(1)}$, where $\Theta^{(1)}$ is defined by Eq.~\eqref{theta1}. Let $\boldsymbol v_i$ be the $i$th row of $T(0)$ such that $\mathrm{vec}(T(0)) = (\boldsymbol v_1,\boldsymbol v_2,\ldots, \boldsymbol v_N)^\top$.}
\textbf{Input:} Matrices $\Theta^{(1)}$ and $M(0)^{-1}$; vector $(\boldsymbol v_1,\boldsymbol v_2,\ldots, \boldsymbol v_N)$; edge $(i_0, j_0)$\\
\textbf{Output:} Matrix $\Delta T$\\
\bigskip
\tcc{Compute $\Delta M \mathrm{vec}(T(0))$}
Initialize $N^2$-dimensional vector $u = 0$\\
\While{$k\in \{1,\ldots,N\} \setminus \{i_0, j_0\}$}{$u_{(i_0-1)N + k} \gets \Theta_{i_0}\cdot \boldsymbol v_k$\\$u_{(j_0-1)N + k} \gets \Theta_{j_0}\cdot \boldsymbol v_k$\\$u_{(k-1)N + i_0} \gets \Theta_{i_0}\cdot \boldsymbol v_k$\\$u_{(k-1)N + j_0} \gets \Theta_{j_0}\cdot \boldsymbol v_k$}
$u_{(i_0-1)N + j_0} \gets \Theta_{i_0}\cdot \boldsymbol v_{j_0} + \Theta_{j_0}\cdot \boldsymbol v_{i_0}$\\
$u_{(j_0-1)N + i_0} \gets \Theta_{i_0}\cdot \boldsymbol v_{j_0} + \Theta_{j_0}\cdot \boldsymbol v_{i_0}$\\
\tcc{$u = (u_1, \ldots, u_{N^2})^\top$ is now equal to $\Delta M \mathrm{vec}(T(0))$}
\bigskip
\tcc{Compute $\mathrm{vec}(\Delta T)$}
\tcc{Multiply $M(0)^{-1}$ and the already calculated $\Delta M \mathrm{vec}(T(0))$}
Initialize $N^2$-dimensional vector $\mathrm{vec}(\Delta T) = 0$\\
\While{$k\in \{1,\ldots,N\} \setminus \{i_0, j_0\}$}{$\mathrm{vec}(\Delta T) \gets \mathrm{vec}(\Delta T) + u_{(i_0-1)N + k} \left( M(0)^{-1} \right)_{(i_0-1)N + k}$\\$\mathrm{vec}(\Delta T) \gets \mathrm{vec}(\Delta T) + u_{(j_0-1)N + k} \left( M(0)^{-1} \right)_{(j_0-1)N + k}$\\$\mathrm{vec}(\Delta T) \gets \mathrm{vec}(\Delta T) + u_{(k-1)N+i_0} \left( M(0)^{-1} \right)_{(k-1)N+i_0}$\\$\mathrm{vec}(\Delta T) \gets \mathrm{vec}(\Delta T) + u_{(k-1)N+j_0} \left( M(0)^{-1} \right)_{(k-1)N+j_0}$} $\mathrm{vec}(\Delta T) \gets \mathrm{vec}(\Delta T) + u_{(i_0-1)N+j_0} \left( M(0)^{-1} \right)_{(i_0-1)N+j_0}$\\
$\mathrm{vec}(\Delta T) \gets \mathrm{vec}(\Delta T) + u_{(j_0-1)N+i_0} \left( M(0)^{-1} \right)_{(j_0-1)N+i_0}$\\
\textbf{Return} $\Delta T$
\end{algorithm}

We now discuss the computational complexity of our perturbation method. Because the inner product of $N$-dimensional vectors has a time complexity of $O(N)$, the first while loop in Algorithm~1 has a complexity of $O(N^2)$. The second while loop computes $\mathrm{vec}(\Delta T)$. Because the scalar multiplication of an $N^2$-dimensional vector requires $O(N^2)$ time, the entire while loop has a time complexity of $O(N^3)$. Therefore, for a single perturbation experiment, one can carry out the entire algorithm in $O(N^3)$ time to obtain the perturbed $\{t_{ij}\}$, and hence $(b/c)^*$. This is considerably smaller than $O(N^{4.75})$ and $O(N^6)$ with the Coppersmith-Winograd algorithm and the standard algorithm, respectively. The entire procedure to determine the single edge to be removed to maximize cooperation with the perturbation theory requires $O(N^3 |E|)$ time in general networks and $O(N^4)$ time for sparse networks.

\section{Data}
\label{dataset}

\begin{figure}[!h]
\begin{center}
 \includegraphics[width=0.92\textwidth]{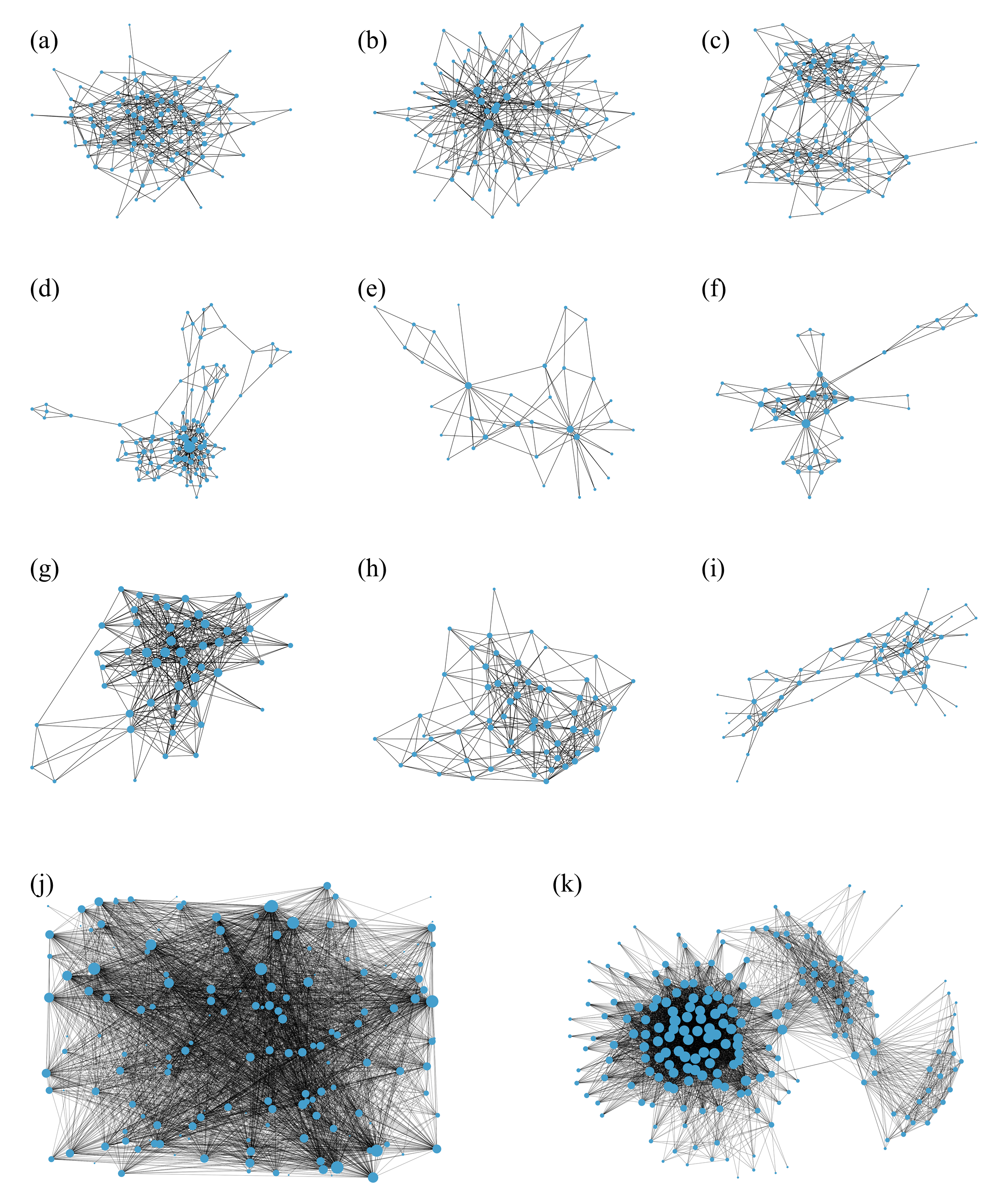}
  \caption{Visualization of the networks used in the numerical analysis. (a) ER random graph. (b) BA model. (c) Planted 2-partition model. (d) LFR model. (e) Karate club. (f) Weaver. (g) Sparrow. (h) Lizard. (i) Dolphin. (j) Email. (k) Bird. All the networks are undirected. The linear size of the node is proportional to its degree. We have ignored the weight of the edge in this figure and our analysis.}
  \label{fig:visualization}
\end{center}
\end{figure}

We use the following four synthetic networks and seven empirical networks in our numerical analysis in section~\ref{results}. We show the number of nodes and that of edges for each network in Table~\ref{tab: network property} and visualize them in Fig.~\ref{fig:visualization}. All the networks are connected networks without self-loops.

We use a network generated by the Erd\H{o}s–R\'{e}nyi (ER) random graph with $N=100$ nodes. We connect $300$ pairs of nodes out of the $N(N-1)/2=4950$ pairs of nodes selected uniformly at random. The average degree $\langle k\rangle = 6$. 

With the Barab\'{a}si–Albert (BA) model, we sequentially add new nodes each with $m=3$ edges that connect to existing nodes according to the linear preferential attachment rule \cite{barabasi1999emergence}. We start the growth process from the star graph with four nodes. The degree distribution approximately obeys $p(k)\propto k^{-3}$, where $p(k)$ is the probability that a node has degree of $k$, and $\propto$ represents ``proportional to'', in the limit of $N\to \infty$. We set $N=100$ and $m=3$, which yields $291$ edges, implying $\langle k\rangle = 5.82 \approx 6$. 

The planted $\ell$-partition model, also called the random partition (RP) graph, partitions the set of $N$ nodes into $\ell$ groups, each of which has $N/\ell$ nodes \cite{fortunato2010community}. Any pair of nodes in the same group is adjacent to each other with probability $p_{\mathrm{in}}$. Any pair of nodes belonging to different groups are adjacent to each other with probability $p_{\mathrm{out}}$. If $p_{\mathrm{in}}>p_{\mathrm{out}}$, the intra-cluster edge density exceeds the inter-cluster edge density such that the network has community structure. We set $N = 100$, $\ell = 2$, $p_{\mathrm{in}}=0.11$, and $p_{\mathrm{out}}=0.01$ such that the mean degree $\langle k\rangle = p_{\mathrm{in}}(N/\ell - 1) + p_{\mathrm{out}}N(\ell-1)/\ell = 5.89$ in theory. We use a network generated by this model having $\langle k\rangle = 6.12$.

The Lancichinetti–Fortunato–Radicchi (LFR) model generates networks with community structure \cite{lancichinetti2008benchmark}. The model generates a power-law degree distribution with power-law exponent $\gamma$, and a power-law distribution of the size of the community with power-law exponent $\kappa$. The model also requires the maximal degree $k_{\mathrm{max}}$ and mean degree $\langle k \rangle$ as input. The mixing parameter $\overline{\mu} \in (0,1)$ specifies the fraction of edges that connect different communities. A small value of $\overline{\mu}$ leads to strong community structure. We set $N=100$, $\gamma=3$, $\kappa=2$, $\langle k \rangle = 6$, $k_{\mathrm{max}}=100$, and $\overline{\mu}=0.1$. A network generated by this model that we use has $\langle k\rangle = 6.08$.

We consider the following seven empirical networks. The karate club network consists of 34 nodes and 78 edges \cite{zachary1977information}. Each node represents a member of a karate club in a university in the United States, who were observed between 1970 and 1972. The edges represent interaction outside the activities of the club. 

The weaver network has 42 nodes and 151 edges \cite{van2014cooperative}. Each node represents a sociable weaver (\textit{Philetairus socius}) observed in Benfontein Game Farm, Kimberley, South Africa. The observation lasted for 10 months in total: September–December 2010 and 2011, and January–February 2013. Two nodes are adjacent to each other if the two weavers used the same nest chambers either for roosting or nest-building within a series of observations in the same year.

The sparrow network has 52 nodes and 516 edges \cite{arnberg2015social}. A node represents a golden-crowned sparrow (\textit{Zonotrichia atricapilla}) observed at the University of California, Santa Cruz Arboretum. The data was recorded between January and March 2010 \cite{arnberg2015social}. Although the original network is weighted, we regard this network as an unweighted network.

The lizard network has 60 nodes and 318 edges \cite{bull2012social}. Each node represents a lizard (\textit{Tiliqua rugosa}) observed in a chenopod shrubland near Bundey Bore Station in South Australia. Each lizard was  attached to the dorsal surface of the tail a data logger unit, which recorded synchronized GPS locations every 10 minutes. Two lizards were regarded to be adjacent to each other if they were within 2 meters of each other in any GPS record.

The dolphin network has 62 nodes and 159 edges \cite{lusseau2003bottlenose}. Each node represents a bottlenose dolphin (\textit{Tursiops}). An edge represents a frequent association between two dolphins. 

The email network has 167 nodes and 3251 edges \cite{michalski2011matching}. Each node represents an employee of a mid-sized manufacturing company in Poland. An edge between two nodes (i.e., employees) indicates that there exists at least one email correspondence between the two individuals. We do not distinguish the senders and the recipients and treat the network as undirected network.

The bird network has 202 nodes and 11900 edges \cite{firth2015experimental}. In the experiment, they placed some nest boxes in Wytham Woods, Oxford, UK, for six days to record individuals that landed on the entrance hole while prospecting for breeding territories. Each node represents a wild bird, which is either great tit (\textit{Parus major}), blue tit (\textit{Cyanistes caeruleus}), marsh tit (\textit{Poecile palustris}), coal tit (\textit{Periparus ater}), or Eurasian nuthatch (\textit{Sitta europaea}). An edge represents two birds that overlapped in nest-box exploration patterns on the same day.

\section{Numerical results}
\label{results}

\subsection{Addition or removal of a single edge\label{sub:numerical-add-remove-1edge}}

We examine the accuracy at which our perturbation theory describes the change in $(b/c)^*$ when we add or remove an edge in the given unweighted network. Before the perturbation, $w_{ij} = w_{ji} = 1$ if there exists an edge between the $i$th and $j$th nodes, and $w_{ij} = w_{ji} = 0$ otherwise. 
In the case of edge addition, we add an edge with weight $\varepsilon$ between a pair of nodes $(i_0, j_0)$ without an edge in the original network unless we state otherwise, where $0 < \varepsilon \leq 1$. Therefore, $w_{i_0 j_0} (= w_{j_0 i_0})$ changes from $0$ to $\varepsilon$, and all the other $w_{ij} \in \{0, 1 \}$ values remain unchanged. The addition of an unweighted edge corresponds to $\varepsilon=1$. In the case of edge removal, we reduce the weight of an edge $(i_0, j_0)$ in the original network by $-\varepsilon$, where $-1 \leq \varepsilon < 0$. Therefore, $w_{i_0 j_0} (= w_{j_0 i_0})$ changes from $1$ to $1 + \varepsilon$, and all the other $w_{ij}$ values remain unchanged. The complete removal of an unweighted edge corresponds to $\varepsilon = -1$.

We are interested in whether the linear approximation to $(b/c)^*(\varepsilon)$ given by Eq.~(\ref{bcthreshold epsilon}), i.e., $\Delta (b/c)^*$, which we call the slope, predicts the change in $(b/c)^*$ in response to the addition of a single edge, i.e., $(b/c)^*(1)-(b/c)^*(0)$, or the removal of a single edge, i.e., $(b/c)^*(-1)-(b/c)^*(0)$. We start by directly computing the change in $(b/c)^*$, i.e., $(b/c)^*(\varepsilon)-(b/c)^*(0)$, in response to adding a new edge of weight $\varepsilon$ $(>0)$ or reducing the weight of an existing edge by changing the edge weight to $1+\varepsilon$ $(<1)$ for various values of $\varepsilon$ for relatively small networks. The outcome of our perturbation theory, i.e., $\Delta (b/c)^*$ is equal to $\lim_{\varepsilon \to 0}\left[ (b/c)^*(\varepsilon) - (b/c)^*(0)\right]/\varepsilon$, where $(b/c)^*(0)$ and $(b/c)^*(\varepsilon)$ are the values obtained by the direct numerical simulations. 

\begin{figure}[!h]
  \includegraphics[width=0.98\textwidth]{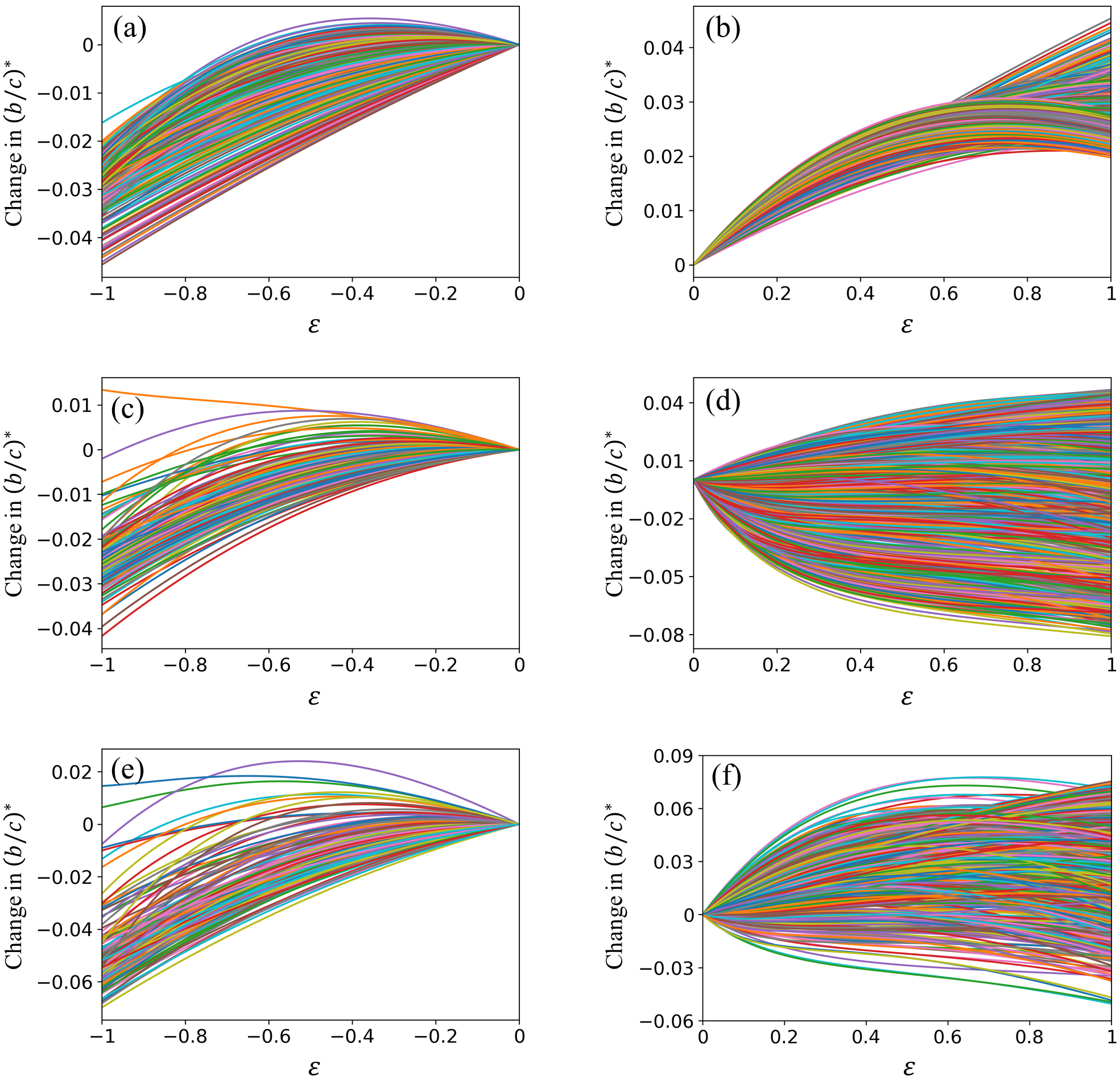}
  \caption{Change in $(b/c)^*$ as a function of the change in the edge weight, $\varepsilon$. (a) BA model, removal of an existing edge. (b) BA model, addition of a new edge. (c) Planted 2-partition model, removal of an existing edge. (d) Planted 2-partition model, addition of a new edge. (e) Dolphin network, removal of an existing edge. (f) Dolphin network, addition of a new edge. In (a), (c), and (e), each line represents an edge in the original network. In (b), (d), and (f), each line represents a pair of nodes that is not adjacent to each other in the original network. The line color is only as a guide to the eyes.}
  \label{Cuver plot}
\end{figure}

We show the relationship between $(b/c)^*(\varepsilon) - (b/c)^*(0)$ and $\varepsilon$ when we reduce the weight of a single edge in a BA network with $N=100$ nodes in Fig.~\ref{Cuver plot}(a). Each line in the figure corresponds to an edge whose weight is gradually reduced. Note that $\varepsilon = 0$ corresponds to the original network. Figure~\ref{Cuver plot}(a) indicates that $(b/c)^*$ roughly monotonically decreases as we gradually decrease the edge weight (i.e., decrease $\varepsilon$ from 0 to negative values) except near $\varepsilon = 0$. For this network, the removal of any single edge (i.e., $\varepsilon=-1$) leads to a decrease in $(b/c)^*$, implying that the edge removal promotes cooperation. However, we note that a small decrease in the weight of an edge in the original network (e.g., $\varepsilon = -0.3$) increases $(b/c)^*$ for some edges, making cooperation more difficult than in the original network. Figure~\ref{Cuver plot}(a) implies that the perturbation theory is not accurate at describing the amount of the change in $(b/c)^*$ upon the edge removal because most of the curves shown in the figures, corresponding to the different edges in the original network, are far from being linear. However, we observe that the curves with the largest values of the slope of the curve at $\varepsilon = 0$ tend to yield the smallest values of $(b/c)^*$ at $\varepsilon = -1$. Therefore, the perturbation theory, which produces the slope value, is expected to be efficient at detecting the edges whose removal yields the largest decrease in $(b/c)^*$.

We show in Fig.~\ref{Cuver plot}(b) the change in $(b/c)^*$ plotted against $\varepsilon$ when we add a new edge with weight $\varepsilon$. Each line corresponds to a pair of nodes between which there is initially no edge. Note that $\varepsilon=1$ corresponds to the addition of an unweighted edge. We find that the addition of any unweighted edge increases $(b/c)^*$, making cooperation difficult. However, in contrast to the case of edge removal, the addition of an unweighted edge (i.e., with edge weight $\varepsilon = 1$) does not necessarily yield the largest change in $(b/c)^*$ among edges of different weights $\varepsilon \in (0, 1]$. Specifically, for many node pairs that are initially not adjacent to each other, adding an edge with an intermediate edge weight (e.g., $\varepsilon \approx 0.7$) maximizes the increase in $(b/c)^*$ (see Fig.~\ref{Cuver plot}(b)). Another observation is that the slope of the curve at $\varepsilon = 0$, corresponding to the perturbation theory, is apparently less predictive of the effect of adding an unweighted edge (i.e., $\varepsilon = 1$). Specifically, Fig.~\ref{Cuver plot}(b) indicates that, even if the slope at $\varepsilon = 0$ is large, $(b/c)^*$ at $\varepsilon = 1$ can be relatively small because $(b/c)^*$ decreases as $\varepsilon$ increases when $\varepsilon$ is close to 1. Furthermore, the curves with the largest slopes at $\varepsilon = 0$ do not yield the largest changes in the $(b/c)^*$ value at $\varepsilon = 1$, which implies that the perturbation theory is expected to be inefficient at predicting the edge addition that makes the cooperation most difficult.

We find similar results for the planted $2$-partition model for the gradual removal of a single edge (see Fig.~\ref{Cuver plot}(c)). A notable difference from the case of the BA model is that there exists one edge whose complete removal increases $(b/c)^*$, making the cooperation difficult. The two nodes forming this edge have degrees $2$ and $9$, which are not outstanding. Furthermore, we have confirmed by running a deterministic approximate modularity maximization algorithm \cite{clauset2004finding}, using function greedy\_modularity\_communities in NetworkX, that these two nodes belong to the same community among the four communities detected. Therefore, this particular edge looks like just a normal edge.

We show in Fig.~\ref{Cuver plot}(d) the dependence of $(b/c)^*$ on $\varepsilon$ when we gradually increase the weight of an edge that is initially absent in the planted $2$-partition network. The slope of the curve at $(b/c)^*$ at $\varepsilon = 0$ is apparently not strongly related to the change in $(b/c)^*$ at $\varepsilon = 1$.

We show the results of edge removal in the dolphin network in Fig.~\ref{Cuver plot}(e). There are two edges out of the 150 edges of which the removal (i.e., $\varepsilon = -1$) increases $(b/c)^*$, making cooperation difficult. These two edges are formed by two nodes with degrees 2 and 5 and two other ones with degrees 2 and 7. These degree values are not outstanding in the entire network. The four nodes belong to the same community among the four communities detected by the same approximate modularity maximization algorithm \cite{clauset2004finding}. These results suggest that the two edges apparently look normal. The removal of any other edge decreases $(b/c)^*$, enhancing cooperation. Similar to the BA model, the curves with the largest slopes at $\varepsilon = 0$ yield the largest decreases in $(b/c)^*$ at $\varepsilon = -1$. 

We show in Fig.~\ref{Cuver plot}(f) the dependence of $(b/c)^*$ on $\varepsilon$ when we gradually increase the weight of an edge that is initially absent in the dolphin network. The results are similar to those for the planted $2$-partition model shown in Fig.~\ref{Cuver plot}(d). Many curves yield decrease in $(b/c)^*$ at $\varepsilon = 1$, implying that the edge addition can promote cooperation, whereas the converse is the case for many other curves. The slope of the curve of $(b/c)^*$ at $\varepsilon = 0$ is apparently not strongly related to the change in $(b/c)^*$ at $\varepsilon = 1$.

\begin{table}[!t]
    \centering
    \caption{Pearson correlation coefficient, $r$, between the shift in $(b/c)^*$ obtained by direct numerical simulations and that predicted by the perturbation theory. We remind that $N$ is the number of nodes and that $|E|$ is the number of edges. A large positive value of $r$ upon edge addition or enhancement implies that the perturbation theory is good at predicting the outcome of adding or enhancing an edge. A large negative value of $r$ upon edge removal implies that the perturbation theory is good at predicting the outcome of removing an edge.}
    \begin{tabular}{>{\centering\arraybackslash}m{1.3cm} >{\centering\arraybackslash}m{0.7cm} >{\centering\arraybackslash}m{0.7cm} >{\centering\arraybackslash}m{1.5cm} >{\centering\arraybackslash}m{1.5cm} >{\centering\arraybackslash}m{1.5cm}}
    \hline
    \multirow{2}{*}{Network} & \multirow{2}{*}{$N$} & \multirow{2}{*}{$|E|$} & $r$, edge & $r$, edge & $r$, edge \\
    & & & addition & removal & enhancement \\
    \hline
    ER & 100 & 300 & $-0.55$ & $-0.87$ & $0.99$\\
    BA & 100 & 291 & $-0.36$ & $-0.86$ & $0.99$\\
    RP & 100 & 306 & $-0.39$ & $-0.80$ & $0.98$\\
    LFR & 100 & 304 & $\phantom{-}0.27$ & $-0.84$ & $0.98$\\
    Karate & 34 & 78 & $\phantom{-}0.35$ & $-0.88$ & $0.99$\\
    Weaver & 42 & 152 & $\phantom{-}0.94$ & $-0.93$ & $0.99$\\
    Sparrow & 52 & 516 & $-0.01$ & $-0.95$ & $1.00$\\
    Lizard & 60 & 318 & $\phantom{-}0.72$ & $-0.93$ & $0.99$\\
    Dolphin & 62 & 159 & $\phantom{-}0.56$ & $-0.72$ & $0.96$\\
    \hline
    \end{tabular}
    \label{tab: network property}
\end{table}

\begin{figure}[!h]
  \centering
  \includegraphics[width=0.92\textwidth]{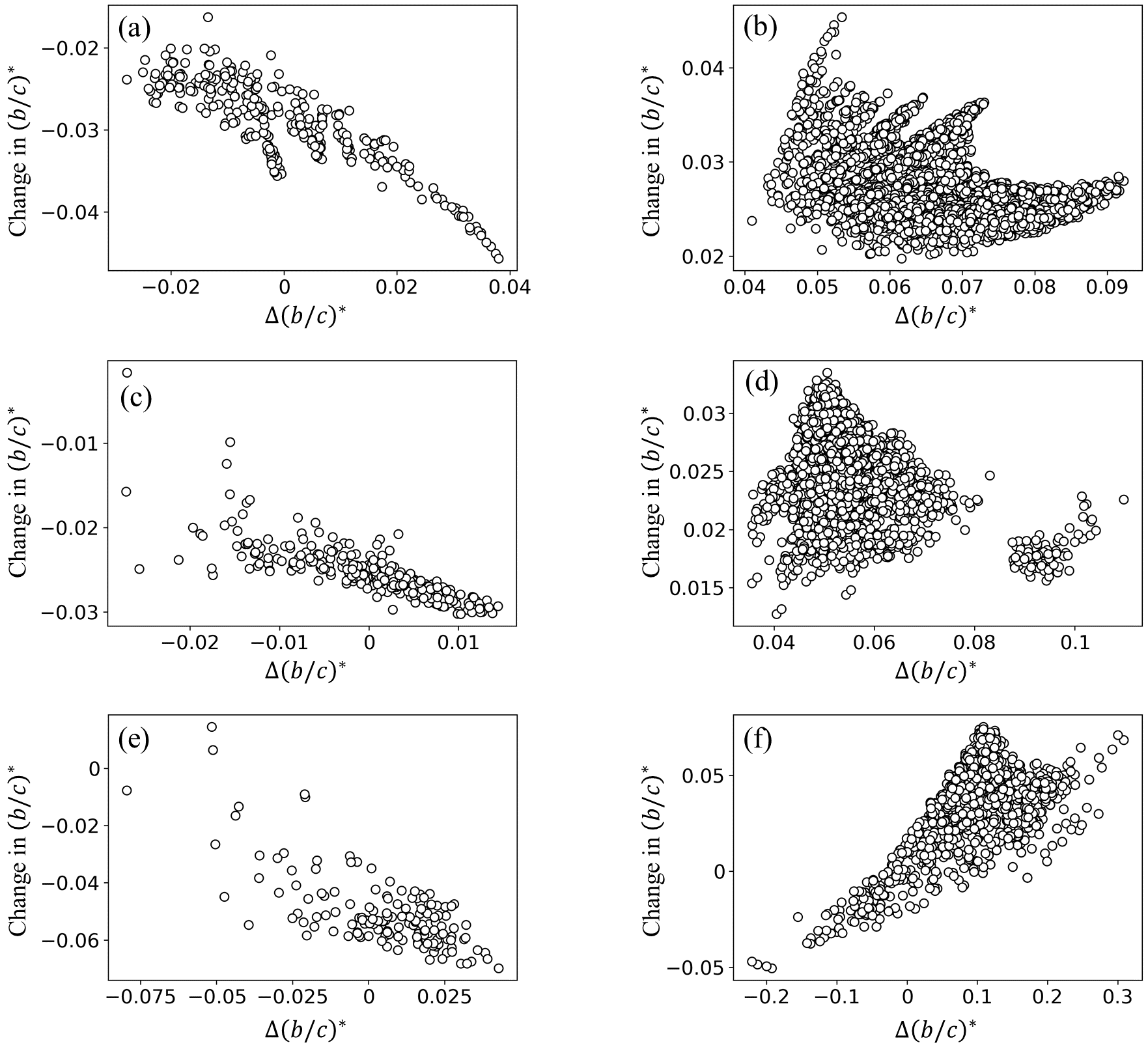}
  \caption{Change in $(b/c)^*$ when we remove or add an unweighted edge as a function of the slope $\Delta (b/c)^*$ of the curves shown in Fig.~\ref{Cuver plot} at $\varepsilon = 0$. (a) BA model, removal of an existing edge. (b) BA model, addition of a new edge. (c) Planted 2-partition model, removal of an existing edge. (d) Planted 2-partition model, addition of a new edge. (e) Dolphin network, removal of an existing edge. (f) Dolphin network, addition of a new edge. Each circle in (a), (c), and (e) represents an edge in the original network. Each circle in (b), (d), and (f) represents a pair of nodes that is not adjacent to each other in the original network.}
  \label{scatter plot}
\end{figure}

The nonlinearity in the curves shown in Fig.~\ref{Cuver plot} indicates that our perturbation theory is not accurate at predicting the amount of change in $(b/c)^*$ when we completely remove or add an edge in most cases. Therefore, we turn to ask whether the slope obtained from the perturbation theory is useful at determining the edge whose removal or addition changes $(b/c)^*$ by a large amount, representing strong promotion or suppression of cooperation in networks. We show in Fig.~\ref{scatter plot}(a) the relationship between the change in $(b/c)^*$ when we remove an edge from the BA network and the slope $\Delta(b/c)^*$ obtained from Eq.~(\ref{bcthreshold epsilon}). The two quantities are strongly negatively correlated (Pearson correlation coefficient $r=-0.86$, sample size $n = 291, p<0.01$). This result indicates that the perturbation theory, which is theoretically accurate only in the vicinity of $\varepsilon = 0$, is good at predicting the outcome of removing an edge. We show in Fig.~\ref{scatter plot}(b) the change in $(b/c)^*$ when we add a new edge to the same BA network as a function of the slope, $\Delta (b/c)^*$. The change in $(b/c)^*$ is not strongly positively correlated with $\Delta (b/c)^*$, suggesting that the perturbation theory is not good at predicting the outcome of adding an edge, whereas the correlation coefficient is
significant due to a large sample size ($r=-0.36, n = 4659, p<0.01$). Note that a large positive correlation coefficient when we add an edge would imply that the perturbation theory is good at predicting the outcome of adding an edge.

We show in Figs.~\ref{scatter plot}(c) and~\ref{scatter plot}(d) the results for the same correlation analysis for the planted $2$-partition model network. When one removes an existing edge, the change in $(b/c)^*$ and slope $\Delta (b/c)^*$ are strongly negatively correlated ($r=-0.80, n = 306, p<0.01$; see Fig.~\ref{scatter plot}(c)), which is similar to the result for the BA model shown in Fig.~\ref{scatter plot}(a), suggesting that the perturbation theory is good at predicting the outcome of removing an edge. When one adds a new edge, the change in $(b/c)^*$ and slope $\Delta (b/c)^*$ are weakly correlated for this network ($r=-0.39, n=4644, p<0.01$; see Fig.~\ref{scatter plot}(d)), which is similar to the result for the BA model shown in Fig.~\ref{scatter plot}(b).

We show the corresponding results for the dolphin network in Figs.~\ref{scatter plot}(e) and~\ref{scatter plot}(f). The change in $(b/c)^*$ and slope $\Delta (b/c)^*$ are strongly negatively correlated when one removes an edge ($r=-0.72, n=150, p<0.01$; see Fig.~\ref{scatter plot}(e)) and less strongly correlated when one adds a new edge ($r=0.56, n=1732, p<0.01$; see Fig.~\ref{scatter plot}(f)). A strongly negative correlation for the edge removal (i.e., $r=-0.72$) is similar to the result for the BA model. A positive correlation for the edge addition (i.e., $r=0.56$) implies that the perturbation theory is to some extent good at predicting the outcome of adding an edge.

We show in Table~\ref{tab: network property} the same relationships for the other networks. For all synthetic and empirical networks, the slope $\Delta (b/c)^*$ obtained from perturbation theory is strongly negatively correlated with the change in $(b/c)^*$ when we remove an existing edge ($r\leq -0.72$). Therefore, the perturbation theory is effective at predicting the outcome of removing an edge across different networks. However, the correlation is strongly positive only for a small fraction of networks (i.e., $r\ge 0.5$ for three out of the nine networks) when we add a new edge to the network. 

\subsection{Enhancement of the weight of an existing edge}

In this section, we allow weighted networks and consider an increase or decrease in the weight of an existing edge of the network. Because we effectively analyzed the case of the decrease in the edge weight in
section~\ref{sub:numerical-add-remove-1edge} (i.e., by setting $-1 < \varepsilon < 0$), here we only consider enhancement of the weight of an existing edge by $\varepsilon$.

\begin{figure}[!h]
  \centering
  \includegraphics[width=0.92\textwidth]{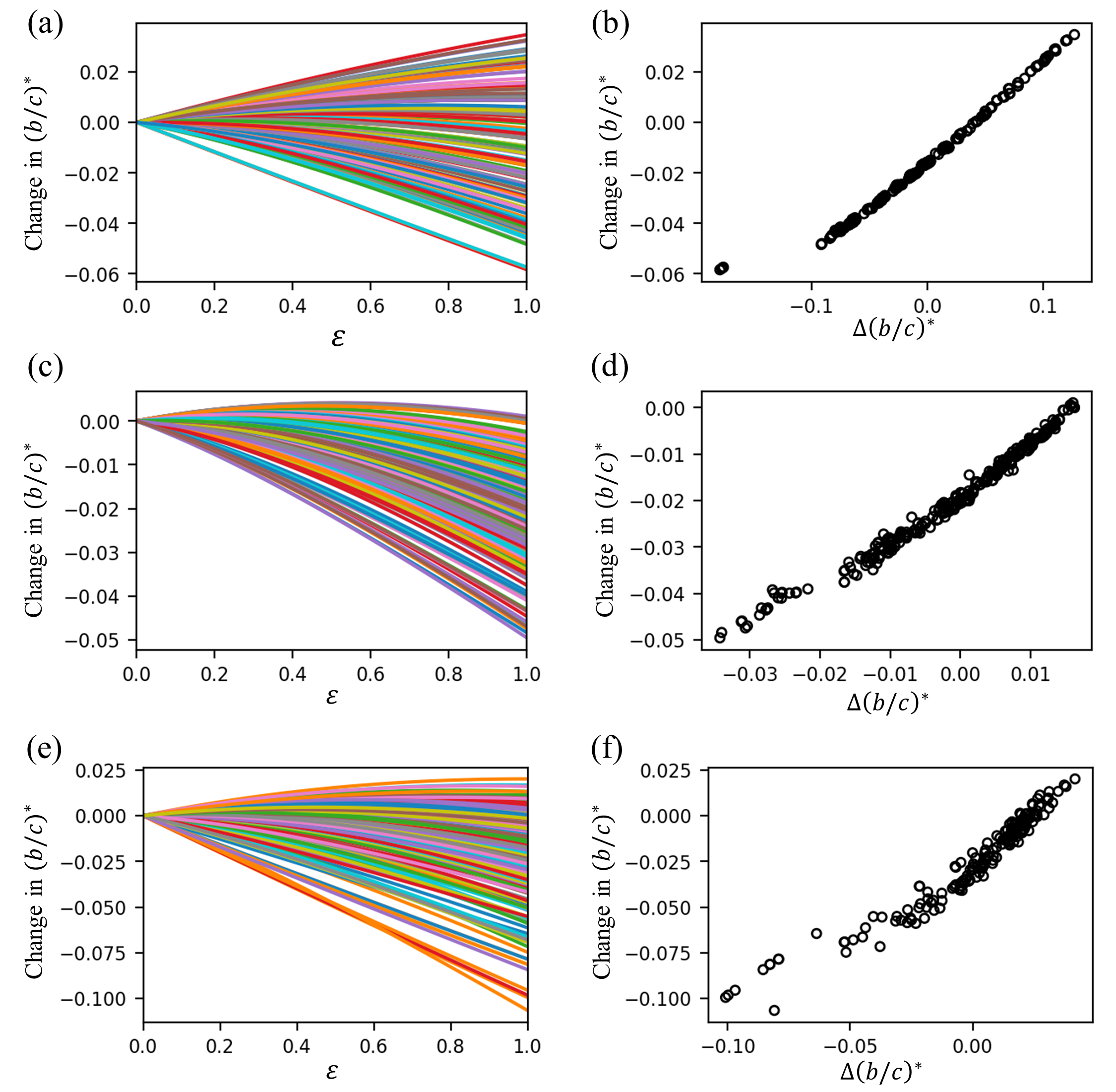}
  \caption{Change in $(b/c)^*$ when one enhances the weight of an existing edge. Panels (a), (c), (e): Change in $(b/c)^*$ as a function of the increase in the weight of an existing edge, $\varepsilon$. Each line represents an edge in the original network. Panels (b), (d), (f): Change in $(b/c)^*$ when we enhance an edge weight by $\varepsilon=1$, plotted against the slope $\Delta (b/c)^*$ of the curves shown in panels (a), (c), and (e) at $\varepsilon = 0$. Each circle represents an edge in the original network. (a) and (b): BA model. (c) and (d): Planted 2-partition model. (e) and (f): Dolphin network.}
  \label{fig: edge_enhancement}
\end{figure}

We enhanced the weight of an existing edge by $0<\varepsilon \le 1$, making the edge weight $1+\varepsilon$, and numerically examined $(b/c)^*$ in the altered weighted networks. We plot in Figs.~\ref{fig: edge_enhancement}(a), \ref{fig: edge_enhancement}(c), and \ref{fig: edge_enhancement}(e) the change in $(b/c)^*$ relative to the original network against $\varepsilon$ for the three networks used in Figs.~\ref{Cuver plot} and \ref{scatter plot}. For the BA model, increasing the weight of 74 out of the 291 existing edges from $1$ to $2$ (i.e., $\varepsilon = 1$) led to an increase in $(b/c)^*$, making cooperation more difficult, whereas the opposite is the case when one enhances the weight of any other edge (see Fig.~\ref{fig: edge_enhancement}(a)). This result contrasts with the case of adding a new edge to the same network, which always increases $(b/c)^*$ (see Fig.~\ref{Cuver plot}(b)). In the planted 2-partition model (Fig.~\ref{fig: edge_enhancement}(c)) and the dolphin network (Fig.~\ref{fig: edge_enhancement}(e)), enhancing the edge weight of 7 out of the 306 edges and 23 out of the 159 edges, respectively, led to an increase in $(b/c)^*$. Therefore, in a majority of cases, cooperation becomes easier by enhancing the weight of a single edge, which contrasts with the results for adding a new edge to these networks (see Figs.~\ref{Cuver plot}(d) and \ref{Cuver plot}(f)). These results altogether suggest that adding new edges and enhancing the weight of existing edges often lead to different results.

Figures~\ref{fig: edge_enhancement}(a), \ref{fig: edge_enhancement}(c), and \ref{fig: edge_enhancement}(e) also indicate that the change in $(b/c)^*$ is close to linear as a function of $\varepsilon$. Therefore, our perturbation theory should be accurate at estimating the change in $(b/c)^*$ with $\varepsilon=1$. To verify this prediction, we show in Figs.~\ref{fig: edge_enhancement}(b), \ref{fig: edge_enhancement}(d), and \ref{fig: edge_enhancement}(f) the relationship between the change in $(b/c)^*$ in response to changing the weight of a single edge from $1$ to $2$ and $\Delta (b/c)^*$ obtained by the perturbation theory for the three networks. As expected, the accuracy of the perturbation theory is high. We have confirmed that a high accuracy also holds true for other networks (see the last column of Table~\ref{tab: network property}). These high accuracy results are in stark contrast to the results in case of adding a new edge, with which the accuracy of the perturbation theory is low.

\subsection{Sequential edge removal}

The nonlinearity in the curves shown in Fig.~\ref{Cuver plot}, and the results shown in Fig.~\ref{scatter plot} and Table~\ref{tab: network property} indicate that our perturbation theory is not accurate at estimating the amount of change in $(b/c)^*$ upon an edge removal. Therefore, we turn to investigate whether our perturbation theory is good at finding edges to be sequentially removed to decrease $(b/c)^*$ by a large amount in larger networks. Denote by $G_0$ an original network. We remove the edge with the largest $\Delta (b/c)^*$, resulting in network $G_1$. Then, we calculate $\Delta (b/c)^*$ for each existing edge in $G_1$ and remove the edge with the largest $\Delta (b/c)^*$, resulting in network $G_2$. We repeat this procedure another three times to eventually obtain network $G_5$, which has five fewer edges than $G_0$. 

A simple rule of thumb to determine edges to be removed to enhance cooperation is to use the degree of nodes composing the edge. In particular, $(b/c)^*$ for the death-birth rule is small for random regular graphs with small degrees \cite{ohtsuki2006simple} and general networks with a small mean degree \cite{allen2017evolutionary}. Therefore, we test the performance of our perturbation theory against a degree-based heuristic to remove an edge for enhancing cooperation, which we define as follows. Denote by $(i, j)$ the edge to be removed and by $k_i$ and $k_j$ the degree of the $i$th and $j$th nodes, respectively. Note that $k_i = \sum_{\ell=1}^N w_{i\ell} (= \sum_{\ell=1}^N w_{\ell i})$ for our networks, which are unweighted. For each network, we remove the edge whose $k_i+k_j$ is largest. After removing an edge according to this criterion, we select the edge with the largest $k_i + k_j$ in the reduced network and remove it. We repeat this procedure another three times to remove five edges in total. In our numerical experiments described below, we have verified that the selected edges are always the same if the score for the edge is defined by $k_i k_j$ instead of $k_i + k_j$.

We carry out sequential edge removal experiments on three synthetic networks and three empirical networks. Note that the six networks are mostly larger than those used in the previous numerical simulations. For these networks, it is computationally difficult to exactly calculate $(b/c)^*$ for all possible networks with, for example, one edge being removed from the original network.

\begin{figure}[!h]
  \centering
  \includegraphics[width=0.92\textwidth]{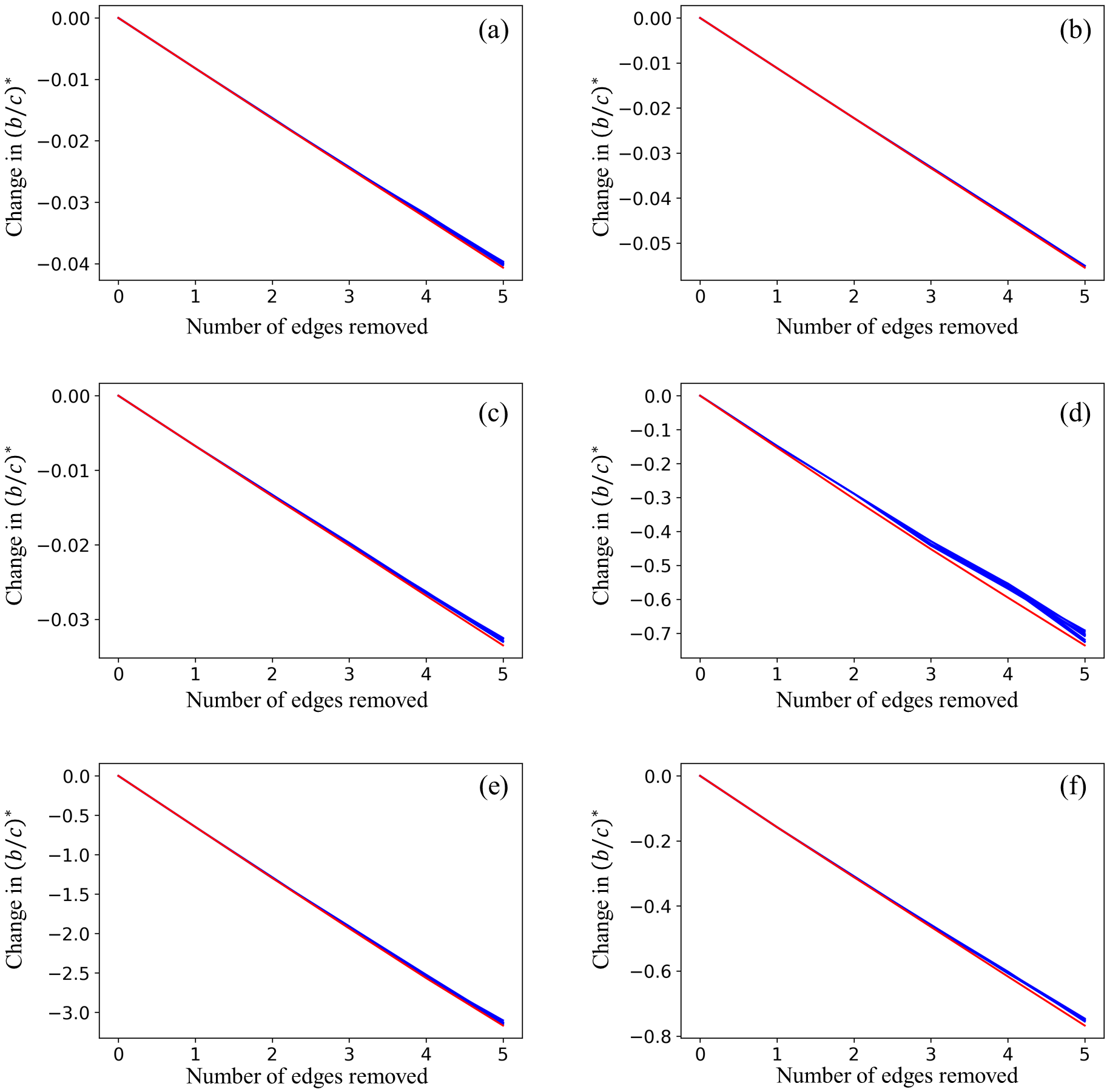}
  \caption{Changes in $(b/c)^*$ upon sequential removal of five edges. (a) ER random graph with 300 nodes and 900 edges. (b) BA model network with 300 nodes and 891 edges. (c) Planted 2-partition network with 300 nodes and 939 edges. (d) Lizard network with 60 nodes and 318 edges. (e) Email network with 167 nodes and 3251 edges. (f) Bird network with 202 nodes and 11900 edges. The red lines represent the edge removal according to the perturbation theory. The blue lines represent the edge removal according to the rank of the degree sum.}
  \label{method comparison}
\end{figure}

We show the change in $(b/c)^*$ relative to the original network as we sequentially remove five edges using our perturbation theory by the red lines in Fig.~\ref{method comparison}. As expected, $(b/c)^*$ decreases, corresponding to negative $\Delta (b/c)^*$ values, as we remove edges one by one. We also show the result of the sequential edge removal based on the degree sum $k_i+k_j$ by the blue lines in the same figure. For all networks, there are multiple edges that have the same value of $k_i + k_j$ at least in one of the five steps to remove a single edge. In this case, we calculated $\Delta (b/c)^*$ for all the possible scenarios of removing one of the edges that maximize $k_i + k_j$ in each step of edge removal. This is why we have obtained multiple blue lines in the figure. In all cases, $(b/c)^*$ decreases as we sequentially remove edges with the largest $k_i + k_j$ value. Figure~\ref{method comparison} indicates that the edge removal based on our perturbation theory results in a larger decrease in $(b/c)^*$ than that based on $k_i + k_j$ for all the networks. To be quantitative, we measured the decrease in $(b/c)^*$ after the removal of five edges compared to the original network with the perturbation theory and with the degree sum. The former was larger than the average of the latter (i.e., average of the blue lines in Fig.~\ref{method comparison}) by a factor of $1.02$, $1.01$, $1.02$, $1.05$, $1.02$, and $1.02$ for the ER random graph (Fig.~\ref{method comparison}(a)), BA model (Fig.~\ref{method comparison}(b)), planted 2-partition network (Fig.~\ref{method comparison}(c)), lizard network (Fig.~\ref{method comparison}(d)), email network (Fig.~\ref{method comparison}(e)), and bird network (Fig.~\ref{method comparison}(f)), respectively.

\section{Conclusions}

To determine $(b/c)^*$ for an arbitrary network, one needs to solve a system of $N^2$ linear equations such that the time complexity is $O(N^6)$. With the Coppersmith-Winograd algorithm, the time complexity is reduced to $O(N^{4.75})$, but this is still large (see section~\ref{algorithm}). In particular, it is computationally costly to carry out graph surgery with various possible edges to be added or removed to compare the results in terms of $(b/c)^*$. Therefore, we have developed a perturbation theory for the graph surgery with which we can evaluate the perturbed $(b/c)^*$ in $O(N^3)$ time. We have verified that the first-order term $\Delta (b/c)^*$ obtained from our perturbation theory predicts the rank of the change in $(b/c)^*$ when one removes an edge from the network with a high accuracy. Specifically, we have numerically shown that the edge with the largest $\Delta (b/c)^*$ value is the one whose actual removal decreases $(b/c)^*$ by the largest amount in two out of the three networks (see Fig.~\ref{scatter plot}(a), \ref{scatter plot}(c), and \ref{scatter plot}(e)). Therefore, we conclude that our perturbation theory is useful for finding the edge whose removal efficiently enhances cooperation in the given network with a reduced computational cost.

We focused on the death-birth process because it tends to foster cooperation compared to other rules of strategy updating \cite{ohtsuki2006simple, szabo2007evolutionary}. However, it is straightforward to formulate similar perturbation methods in the case of other updating rules such as the birth-death process \cite{lieberman2005evolutionary, ohtsuki2006simple} and the pairwise comparison rule \cite{blume1993statistical, szabo1998evolutionary, nowak2004emergence, traulsen2006stochastic} as well as in the case of other payoff matrices. In particular, our theory should be applicable to the case of constant selection \cite{lieberman2005evolutionary, allen2021fixation}, with which the payoff matrix is independent of the opponent's action. The perturbation theory may be more accurate for other update rules or games than the combination of the death-birth rule and the prisoner's dilemma game examined in the present study. Exploitation of our perturbation approach in these directions is left for future work.

Another direction of future work is interaction between the selection strength and network perturbation. In the present work, we have assumed the weak selection limit. However, one can retain a selection strength parameter (which is $\eta$ in this article) to be finite and write down a formal solution. Then, it may be interesting to consider the simultaneous limit of weak selection $\eta \to 0$ and weak network perturbation $\varepsilon \to 0$ in a way $\eta$ and $\varepsilon$ are interrelated. Apart from this research direction, assessing the validity of the present perturbation theory under strong selection is left for future work. To this end, we first need to understand the accuracy of the original theory
of fixation of cooperation in networks~\cite{allen2017evolutionary}, which our theory is based on, under strong selection.

We do not know why the perturbation theory is more accurate when one removes an edge than when one adds an edge. Furthermore, we have found that the perturbation theory is fairly accurate at predicting the result for adding a parallel edge where an edge already exists, whereas it is not accurate when adding an edge where an edge does not exist in the original network. In a related vein, we observed nonmonotonic behavior in the cooperativity in terms of $(b/c)^*$ especially when we gradually added a weighted edge (Figs.~\ref{Cuver plot}(b) and~\ref{Cuver plot}(f)). These results lead us to hypothesize that we can engineer networks that promote cooperation better by considering weighted networks than unweighted networks. These topics also warrant future work.

\section*{Data accessibility}
The empirical network data sets are open resources and available at \cite{karate, weaver, sparrow, lizard, dolphin, email, bird}. The Python codes and synthetic network data sets used in the present study are available on GitHub \cite{code}.

\section*{Acknowledgments}
N.M. acknowledges support from AFOSR European Office (under Grant No. FA9550-19-1-7024), the Sumitomo Foundation, the Japan Science and Technology Agency (JST) Moonshot R\&D (under Grant No. JPMJMS2021), and the National Science Foundation (under Grant Nos. 2052720 and 2204936).

\begin{appendices}

\section{Derivation of Eq.~\eqref{eq_delta_pi}}\label{sec:derivation-Delta_pi}

We restate the stationary probability $\pi_i(0)$, given by Eq.~\eqref{eq:pi_i}, as
\begin{align}
    \pi_i(0) = \frac{s_i(0)}{S(0)},
\end{align}
where $s_i(0)$ is the original weighted degree of the $i$th node, and $S(0)$, given by 
Eq.~\eqref{network weight}, is the original total edge weight of the network. If one perturbs the weight of an edge $(i_0, j_0)$ by $\varepsilon$, the weighted degree of the $i$th node, $s_i(\varepsilon)$, is equal to $s_i(0)+\delta_{ii_0}\varepsilon+\delta_{ij_0}\varepsilon$. After the perturbation, the total edge weight of the network changes to $S(\varepsilon)=S(0)+2\varepsilon$. Therefore, the stationary probability after the perturbation, $\pi_i(\varepsilon)$, is given by
\begin{align}    \pi_i(\varepsilon)=\frac{s_i(0)+\delta_{ii_0}\varepsilon+\delta_{ij_0}\varepsilon}{S(0)+2\varepsilon}.
\end{align}
The first-order derivative of $\pi_i(\varepsilon)$ in terms of $\varepsilon$, denoted by $\Delta \pi_i$, is given by
\begin{align}
\Delta \pi_i(\varepsilon) = \frac{(\delta_{ii_0}+\delta_{ij_0})\left[S(0)+2\varepsilon\right]-\left[s_i(0)+\delta_{ii_0}\varepsilon+\delta_{ij_0}\varepsilon\right]\times 2}{\left[S(0)+2\varepsilon\right]^2}.
\label{eq:Taylor-coef-pi-1}
\end{align}
By setting $\varepsilon=0$ in Eq.~\eqref{eq:Taylor-coef-pi-1}, we obtain
\begin{align}
    \Delta \pi_i(0) & = \frac{S(0)(\delta_{ii_0}+\delta_{ij_0})-2s_i(0)}{S(0)^2} \notag\\
    & = 
    %
    \frac{\delta_{ii_0}+\delta_{ij_0}}{S(0)} - \frac{2\pi_i(0)}{S(0)}.
\end{align}
We wrote $S$ instead of $S(0)$ in Eq.~\eqref{eq_delta_pi} for brevity.

\section{Derivation of Eqs.~\eqref{theta1}, \eqref{theta2}, and \eqref{theta3}}\label{sec:derivation thetas}

\subsection{Derivation of Eq.~\eqref{theta1}}

Let $p_{ij}(\varepsilon)$ be the transition probability of the random walk from node $i$ to node $j$ on the network after perturbing the weight of edge $(i_0, j_0)$ by $\varepsilon$. Note that $p_{ij}(0)$ is the transition probability on the original network. By definition, we have
\begin{align}
p_{ij}(\varepsilon) = \frac{w_{ij}(\varepsilon)}{s_i(\varepsilon)},
\label{eq:p_{ij}(epsilon)}
\end{align}
where $w_{ij}(\varepsilon)$ represents the weight of edge $(i, j)$ after the same perturbation, and we remind that $s_i(\varepsilon)$ is the weighted degree of the $i$th node after the perturbation. In Appendix~\ref{sec:derivation-Delta_pi}, we showed that
\begin{align}
s_i(\varepsilon)=s_i(0)+\delta_{ii_0}\varepsilon+\delta_{ij_0}\varepsilon.
\label{eq:s_i(epsilon)}
\end{align}
We can also verify that
\begin{align}
w_{ij}(\varepsilon)=w_{ij}(0)+\varepsilon\chi_{i_0j_0}(i, j),
\label{eq:w_{ij}(epsilon)}
\end{align}
where $\chi_{i_0j_0}(i, j)$, defined by Eq.~\eqref{chi}, is the indicator function, which is equal to $1$ if and only if edge $(i, j)$ is the perturbed edge $(i_0, j_0)$; it is equal to $0$ otherwise. By substituting Eqs.~\eqref{eq:s_i(epsilon)} and \eqref{eq:w_{ij}(epsilon)} in Eq.~\eqref{eq:p_{ij}(epsilon)}, we obtain
\begin{align}
p_{ij}(\varepsilon) = \frac{w_{ij}(0)+ \varepsilon\chi_{i_0j_0}(i, j)}{s_i(0)+\delta_{ii_0}\varepsilon+\delta_{ij_0}\varepsilon}.
\label{eq:p_{ij}(epsilon)-detail}
\end{align}

By taking the first-order derivative of Eq.~\eqref{eq:p_{ij}(epsilon)-detail} with respect to $\varepsilon$, we obtain
\begin{align}
p_{ij}^\prime (\varepsilon) = \frac{\chi_{i_0j_0}(i, j)\left[s_i(0)+\delta_{ii_0}\varepsilon+\delta_{ij_0}\varepsilon\right] - \left[w_{ij}(0)+ \varepsilon\chi_{i_0j_0}(i, j)\right](\delta_{ii_0}+\delta_{ij_0})}{\left[s_i(0)+\delta_{ii_0}\varepsilon+\delta_{ij_0}\varepsilon\right]^2},
\label{eq:p'_{ij}(epsilon)}
\end{align}
which leads to
\begin{align}
    p_{ij}^\prime (0) & = \frac{\chi_{i_0j_0}(i, j)s_i(0) - w_{ij}(0)(\delta_{ii_0}+\delta_{ij_0})}{\left[s_i(0)\right]^2} \nonumber\\
    & = \frac{\chi_{i_0j_0}(i, j)}{s_i} - p_{ij}(0)\frac{\delta_{ii_0}+\delta_{ij_0}}{s_i}.
\label{eq:p_{ij}'(0)}
\end{align}
We omitted the argument of $s_i(0)$ in the last line of Eq.~\eqref{eq:p_{ij}'(0)} without ambiguity.
This completes the proof of Eq.~\eqref{theta1}. Note that $p_{ij}^\prime (0)$ is equivalent to $\theta^{(1)}_{ij}$ in Eq.~\eqref{theta1}. We chose a different notation $\Theta^{(1)}$ in the main text to avoid confusion between the power of matrix $P$ and the derivative of $P$.

\subsection{Derivation of Eq.~\eqref{theta2}}

We have shown that
\begin{align}
    P(\varepsilon) = P(0) + \varepsilon \Theta^{(1)} + o(\varepsilon),
    \label{appendix:p}
\end{align}
where $\Theta^{(1)} = (\theta^{(1)}_{ij}) = (p_{ij}^\prime (0))$. By squaring Eq.~\eqref{appendix:p}, we obtain
\begin{align}
    P^2(\varepsilon) = P^2(0) + \varepsilon \left[\Theta^{(1)} P(0) + P(0) \Theta^{(1)}\right] + o(\varepsilon).
\end{align}

Let us evaluate matrix $\Theta^{(1)} P(0)$ first. We denote by $a_{ij}$ the $(i, j)$ entry of $\Theta^{(1)} P(0)$. By substituting Eq.~\eqref{theta1} in 
\begin{align}
    a_{ij} = \sum_{k=1}^N \theta^{(1)}_{ik} p_{kj}(0),
    \label{aij}
\end{align}
we obtain
\begin{align}
    a_{ij} & = \sum_{k=1}^N \left[ \frac{\chi_{i_0j_0}(i, k)}{s_i} - p_{ik}(0)\frac{\delta_{ii_0}+\delta_{ij_0}}{s_i} \right] p_{kj}(0) \notag\\
    & = \sum_{k=1}^N \frac{\chi_{i_0j_0}(i, k) p_{kj}(0)}{s_i} - \sum_{k=1}^N p_{ik}(0)p_{kj}(0)\frac{\delta_{ii_0}+\delta_{ij_0}}{s_i} \notag\\
    & = \sum_{k=1}^N \frac{\chi_{i_0j_0}(i, k) p_{kj}(0)}{s_i} - p_{ij}^{(2)}(0)\frac{\delta_{ii_0}+\delta_{ij_0}}{s_i},
    \label{appendix:aij derivation}
\end{align}
where $p_{ij}^{(2)}(0)$ is the transition probability of the random walk from node $i$ to node $j$ in two time steps. Note that $\chi_{i_0j_0}(i, k)=1$ if and only if $(i, k)=(i_0, j_0)$ or $(i, k)=(j_0, i_0)$. If $i=i_0$, then $k=j_0$ must hold true for the term $\chi_{i_0j_0}(i, k) p_{kj}(0)$ not to vanish. Similarly, if $i=j_0$, then $k=i_0$ must hold true for $\chi_{i_0j_0}(i, k) p_{kj}(0)$ not to vanish. If $i \neq i_0, j_0$, then we obtain $\chi_{i_0j_0}(i, k) p_{kj}(0) = 0$ $\forall k$. Therefore, we can simplify Eq.~\eqref{appendix:aij derivation} into
\begin{align}
    a_{ij} & = \frac{\delta_{ii_0}p_{j_0j}(0)}{s_i} + \frac{\delta_{ij_0}p_{i_0j}(0)}{s_i} - p_{ij}^{(2)}(0)\frac{\delta_{ii_0}+\delta_{ij_0}}{s_i}.
    \label{appendix: aij final}
\end{align}

Similarly, by substituting Eq.~\eqref{theta1} in 
\begin{align}
    b_{ij} \equiv \sum_{k=1}^N p_{ik}(0) \theta^{(1)}_{kj},
    \label{bij}
\end{align}
we obtain
\begin{align}
    b_{ij} & = \sum_{k=1}^N p_{ik}(0) \left[ \frac{\chi_{i_0j_0}(k, j)}{s_k} - p_{kj}(0)\frac{\delta_{ki_0}+\delta_{kj_0}}{s_k} \right] \notag\\
    & = \sum_{k=1}^N \frac{\chi_{i_0j_0}(k, j)p_{ik}(0)}{s_k} - \sum_{k=1}^N p_{ik}(0)p_{kj}(0)\frac{\delta_{ki_0}+\delta_{kj_0}}{s_k} \notag\\
    %
    & = \frac{\delta_{jj_0}p_{ii_0}(0)}{s_{i_0}} + \frac{\delta_{ji_0}p_{ij_0}(0)}{s_{j_0}} - p_{ii_0}(0)p_{i_0j}(0) \frac{1}{s_{i_0}} - p_{ij_0}(0)p_{j_0j}(0) \frac{1}{s_{j_0}}.
    \label{appendix: bij final}
\end{align}
Because $\theta^{(2)}_{ij} = a_{ij} + b_{ij}$, we have proved Eq.~\eqref{theta2}.

\subsection{Derivation of Eq.~\eqref{theta3}}

By matrix multiplication, we obtain
\begin{align}
    \Theta^{(3)} = \Theta^{(1)} P^2(0) + P(0) \Theta^{(1)} P(0) + P^2(0) \Theta^{(1)}.
    \label{Theta3}
\end{align}
Let $c_{ij}$, $d_{ij}$, and $e_{ij}$ be the $(i, j)$ entry of matrices $\Theta^{(1)} P^2(0)$, $P(0) \Theta^{(1)} P(0)$, and $P^2(0) \Theta^{(1)}$, respectively.

Using Eq.~\eqref{theta1}, we obtain
\begin{align}
    c_{ij} &= \sum_{k=1}^N \theta^{(1)}_{ik} p^{(2)}_{kj}(0) \notag\\
    & = \sum_{k=1}^N \left[ \frac{\chi_{i_0j_0}(i, k)}{s_i} - p_{ik}(0)\frac{\delta_{ii_0}+\delta_{ij_0}}{s_i} \right] p^{(2)}_{kj}(0) \notag\\
    & = \sum_{k=1}^N \frac{\chi_{i_0j_0}(i, k)p^{(2)}_{kj}(0)}{s_i} - \sum_{k=1}^N p_{ik}(0)p^{(2)}_{kj}(0)\frac{\delta_{ii_0}+\delta_{ij_0}}{s_i} \notag\\
    & = \frac{\delta_{ii_0}}{s_{i_0}}p^{(2)}_{j_0j}(0) + \frac{\delta_{ij_0}}{s_{j_0}}p^{(2)}_{i_0j}(0) - p^{(3)}_{ij}(0)\frac{\delta_{ii_0}+\delta_{ij_0}}{s_i}.
\end{align}
Similarly, we obtain
\begin{align}
    d_{ij} & = \sum_{k=1}^{N} b_{ik} p_{kj}(0)\nonumber\\
    & = \sum_{k=1}^{N} \left[\frac{\delta_{kj_0}p_{ii_0}(0)}{s_{i_0}} + \frac{\delta_{ki_0}p_{ij_0}(0)}{s_{j_0}} - p_{ii_0}(0)p_{i_0k}(0) \frac{1}{s_{i_0}} - p_{ij_0}(0)p_{j_0k}(0) \frac{1}{s_{j_0}} \right] p_{kj}(0) \nonumber\\
    & = \frac{p_{ii_0}(0)p_{j_0j}(0)}{s_{i_0}} + \frac{p_{ij_0}(0)p_{i_0j}(0)}{s_{j_0}} - \frac{p_{ii_0}(0)p_{i_0j}^{(2)}(0)}{s_{i_0}} - \frac{p_{ij_0}(0)p_{j_0j}^{(2)}(0)}{s_{j_0}}
\end{align}
and 
\begin{align}
e_{ij} & = \sum_{k=1}^{N} p_{ik}^{(2)}(0)\theta_{kj}^{(1)}\nonumber\\
& = \sum_{k=1}^{N} p_{ik}^{(2)}(0) \left[ \frac{\chi_{i_0j_0}(k, j)}{s_k} - p_{kj}(0)\frac{\delta_{ki_0}+\delta_{kj_0}}{s_k} \right]\nonumber\\
& = \frac{\delta_{jj_0}}{s_{i_0}}p^{(2)}_{ii_0}(0) + \frac{\delta_{ji_0}}{s_{j_0}}p^{(2)}_{ij_0}(0) - p^{(2)}_{ii_0}(0)p_{i_0j}(0) \frac{1}{s_{i_0}} - p^{(2)}_{ij_0}(0)p_{j_0j}(0) \frac{1}{s_{j_0}}.
\end{align}
Because $\theta^{(3)}_{ij} = c_{ij}+d_{ij}+e_{ij}$, we have proved Eq.~\eqref{theta3}.

\section{Derivation of $\Delta M$}\label{sec:derivation-DeltaM}

In this section, we derive $\Delta M$ as a block matrix
\begin{align}
\Delta M = 
    \begin{pmatrix}
    \Delta_{11} & \Delta_{12} & \cdots & \Delta_{1N}\\
    \Delta_{21} & \Delta_{22} & \cdots & \Delta_{2N}\\
    \vdots & \vdots & \ddots & \vdots\\
    \Delta_{N1} & \Delta_{N2} & \cdots & \Delta_{NN}
    \end{pmatrix},
\end{align}
where each $\Delta_{ij}$ is an $N\times N$ matrix.

The $i$th row of the diagonal block $\Delta_{ii}$ is filled by 0, and all the other rows are the same as those of matrix $-\frac{1}{2}\Theta^{(1)}$. For example, we obtain
\begin{align}
    \Delta_{22}=-\frac{1}{2}
    \begin{pmatrix}
    \theta^{(1)}_{11} & \theta^{(1)}_{12} &  \cdots & \theta^{(1)}_{1N}\\
    0 & 0 & \cdots & 0\\
    \theta^{(1)}_{31} & \theta^{(1)}_{32} & \cdots & \theta^{(1)}_{3N}\\
    \vdots & \vdots & \ddots & \vdots\\
    \theta^{(1)}_{N1} & \theta^{(1)}_{N2} & \cdots & \theta^{(1)}_{NN}
    \end{pmatrix}.
\end{align}
For $i\neq j$, the $j$th row of $\Delta_{ij}$ is equal to the $i$th row of $-\frac{1}{2}\Theta^{(1)}$, and all the other rows are filled by 0. For example, we obtain
\begin{align}
    \Delta_{21} = -\frac{1}{2}
    \begin{pmatrix}
    \theta^{(1)}_{21} & \theta^{(1)}_{22} & \cdots & \theta^{(1)}_{2N}\\
    0 & 0 & \cdots & 0\\
    \vdots & \vdots & \ddots & \vdots\\
    0 & 0 & \cdots & 0\\
    \end{pmatrix}
\end{align}
and 
\begin{align}
    \Delta_{23} = -\frac{1}{2}
    \begin{pmatrix}
    0 & 0 & \cdots & 0\\
    0 & 0 & \cdots & 0\\
    \theta^{(1)}_{21} & \theta^{(1)}_{22} & \cdots & \theta^{(1)}_{2N}\\
    0 & 0 & \cdots & 0\\
    \vdots & \vdots & \ddots & \vdots\\
    0 & 0 & \cdots & 0
    \end{pmatrix}.
\end{align}

Equation~\eqref{theta1} implies that only the $i_0$th and $j_0$th rows of $\Theta^{(1)}$ may be nonzero. 
Owing to this property, there are only
$3N-2$ nonzero matrix blocks $\Delta_{ij}$ out of the $N^2$ blocks of $\Delta M$, which we show using an example as follows.
Assume that we perturb edge $(i_0, j_0) = (4, 7)$. Then, the fourth and seventh rows are the only nonzero rows of $\Theta^{(1)}$. Therefore,
$\Delta_{ii}$, where $i \notin \{i_0, j_0\}$, has only the $i_0$th and $j_0$th rows nonzero. Any off-diagonal matrix block $\Delta_{ij}$, where $i\not\in \{ i_0, j_0 \}$ and $j \neq i$, is zero because it has 
$- \left(\theta_{i1}^{(1)}, \ldots, \theta_{iN}^{(1)}\right)/2$ in the $j$th row, this row is zero given that $i \notin \{i_0, j_0\}$, and all the other rows are zero. 
We next consider $\Delta_{i_0 j}$, where $j\in \{1, \ldots, N\}$.
Diagonal block $\Delta_{i_0 i_0}$ has only the $j_0$th row nonzero, which is given by $-\left(
\theta^{(1)}_{j_0 1}, \ldots, \theta^{(1)}_{j_0 N} \right)/2$. Off-diagonal block $\Delta_{i_0 j}$, where $j \neq i_0$, has only the $j$th row nonzero, which is equal to $-\left( \theta^{(1)}_{i_0 1}, \ldots, \theta^{(1)}_{i_0 N} \right)/2$. 
Therefore, all the blocks $\Delta_{i_0 j}$ with $j\in \{1, \ldots, N\}$ are nonzero in general.
Likewise, $\Delta_{j_0 j_0}$ has only the $i_0$th row nonzero, which is given by $-\left( \theta^{(1)}_{i_0 1}, \ldots, \theta^{(1)}_{i_0 N}\right)/2$. Off-diagonal block $\Delta_{j_0 j}$, where $j \neq j_0$, has only the $j$th row nonzero, which is equal to $-\left( \theta^{(1)}_{j_0 1}, \ldots, \theta^{(1)}_{j_0 N} \right)/2$. Therefore, all the blocks $\Delta_{j_0 j}$ with $j \in \{1, \ldots, N\}$ are nonzero in general. This proves that there are $(N-2) + N + N = 3N-2$ nonzero matrix blocks $\Delta_{ij}$.

Furthermore, most rows of $\Delta M$ are zero rows. Specifically, consider $N$ rows of $\Delta M$ given in a block matrix form by $\left( \Delta_{i1}, \ldots, \Delta_{iN} \right)$, where $i \notin \{ i_0, j_0 \}$. As we have shown, the only non-zero block among $\Delta_{i1}$, $\ldots$, $\Delta_{iN}$ is $\Delta_{ii}$, and the only nonzero rows of $\Delta_{ii}$ are the $i_0$th and $j_0$th rows. Therefore, the $N-2$ rows of $\left( \Delta_{i1}, \ldots, \Delta_{iN} \right)$, i.e., $j$th rows with $j\notin \{ i_0, j_0 \}$, are zero rows. In addition, the $i_0$th row of $\left( \Delta_{i_0 1}, \ldots, \Delta_{i_0 N} \right)$ and the $j_0$th row of $\left( \Delta_{j_0 1}, \ldots, \Delta_{j_0 N} \right)$ are zero rows. Therefore, the number of nonzero rows of $\Delta M$ is equal to $2\times (N-2) + (N-1) \times 2 = 4N-6$.

\end{appendices}

\bibliography{sn-bibliography}
\end{document}